\documentclass[10pt,journal,compsoc]{IEEEtran}
\usepackage[nocompress]{cite}
\usepackage{ragged2e}
\usepackage{amsmath} \interdisplaylinepenalty=2500
\usepackage{microtype,hyphenat,balance,booktabs}
\usepackage{amssymb}
\usepackage{graphicx}
\usepackage[pdftex]{hyperref}
\usepackage{array}
\usepackage{dblfloatfix}
\usepackage[fleqn,tbtags]{mathtools}
\usepackage{flushend}
\usepackage{threeparttable}
\def\smallColSep{\setlength{\arraycolsep}{2pt}}

\newcommand\abs[1]{\left|#1\right|}
\newcommand\norm[1]{\left|\left|#1\right|\right|}
\newcommand{\argmin}{\operatornamewithlimits{argmin}}
\DeclareMathOperator{\Tr}{Tr}

\hypersetup{
    pdfstartview={FitH},    
    pdftitle={Surface Approximation via Asymptotic Optimal Geometric Partition},    
    pdfnewwindow=true,      
    colorlinks=true,       
    linkcolor=blue,          
    citecolor=blue,        
    filecolor=magenta,      
    urlcolor=[rgb]{0.26,0.26,1},           
}

\begin{document}

\title{Surface Approximation via Asymptotic Optimal Geometric Partition}
\author{Yiqi~Cai,
		Xiaohu~Guo,
		Yang~Liu,
		Wenping~Wang,
		Weihua~Mao,
		Zichun~Zhong
		\IEEEcompsocitemizethanks{
		\IEEEcompsocthanksitem Y. Cai and X. Guo are with the Department of Computer Science, The University of Texas at Dallas, Richardson, TX 75083.
		\IEEEcompsocthanksitem Y. Liu is with Microsoft Research Asia, Beijing, China, 100080.
		\IEEEcompsocthanksitem W. Wang is with the Department of Computer Science, The University of Hong Kong, Hong Kong, China.
		\IEEEcompsocthanksitem W. Mao is with the Department of Radiation Oncology, The University of Texas Southwestern Medical Center, Dallas, TX 75390.
		\IEEEcompsocthanksitem Z. Zhong is with the Department of Computer Science, Wayne State University, Detroit, MI 48202.
		\IEEEcompsocthanksitem Corresponding Author: X. Guo, Email: xguo@utdallas.edu}
}
\IEEEtitleabstractindextext{%
\justify
\begin{abstract}	
	In this paper, we present a novel method on surface partition from the perspective of approximation theory. Different from previous shape proxies, the ellipsoidal variance proxy is proposed to penalize the partition results falling into disconnected parts. On its support, the Principle Component Analysis (PCA) based energy is developed for asymptotic cluster aspect ratio and size control. We provide the theoretical explanation on how the minimization of the PCA-based energy leads to the optimal asymptotic behavior for approximation. Moreover, we show the partitions on densely sampled triangular meshes converge to the theoretic expectations. To evaluate the effectiveness of surface approximation, polygonal/triangular surface remeshing results are generated. The experimental results demonstrate the high approximation quality of our method.
\end{abstract}

\begin{IEEEkeywords}
	Surface Approximation, Approximation Theory, Principal Component Analysis, Optimal Asymptotic Behavior, Surface Remeshing.
\end{IEEEkeywords}}

\maketitle
\IEEEdisplaynontitleabstractindextext

%
\IEEEpeerreviewmaketitle

\IEEEraisesectionheading{\section{Introduction}\label{sec:introduction}}
\IEEEPARstart{W}{}ith the ever increasing use of 3D scanners, faithful triangular meshes with dense sampling rate are easy to get for a wide variety of applications. However, for practical purposes, using a huge number of triangles as the surface representation becomes a burden for displaying, storing and processing. Thus, many geometric tasks require a preprocessing step for surface approximation. Depending on the application, this lower-resolution approximation can be triangular representation or non-triangular representation.

The surface approximation problem can be stated as follows: given a dense triangular mesh and a finite polygon/vertex budget, what is the optimal piecewise-linear approximation of the surface under a certain measurement? It has been pointed out in literature to be NP-hard~\cite{Agarwal98}. On the other hand, recent approaches are proposed to generate \emph{asymptotic optimal} results. The term \emph{asymptotic optimal} refers to the fact of exhibiting the maximum efficiency for infinitesimal elements at the asymptotic limit. As shown in approximation theory~\cite{Nadler:86, Simpson:1994:AMT, Chen:07}, the asymptotic optimal results require the alignment and stretching of infinitesimal triangles agreeing with their local curvature tensors.

Instead of explicitly calculating the error-prone curvature tensor, we propose a novel PCA-based energy. This energy also outperforms previous proxy-driven energies by providing cluster size control. We provide the theoretical explanation on how the minimization of the PCA-based energy leads to the optimal asymptotic behavior, as required by approximation theory. Moreover, we show the partitions on dense triangular meshes converge to the theoretic expectations. To evaluate the effectiveness of surface approximation, polygonal/triangular surface remeshing results are generated. The experimental results demonstrate the high approximation quality of our method.

\section{Related Works}
\label{sec::related_works}
\subsection{Geometric Approximation Theory}
Geometric approximation theory is an established mathematical field on how complicated geometry can be best approximated as simpler ones, and with quantitative error measured thereby. 

For surfaces represented as smooth functions, there are some existing works~\cite{Nadler:86, Simpson:1994:AMT, Chen:07} on the asymptotic optimal approximations. For the general $L^p$ metric~\cite{Simpson:1994:AMT}, the optimal piecewise-linear approximation is obtained when each linear element is elongated by $\sqrt{\abs{v_{max}/ v_{min}}}$ along the direction of the maximum eigenvector of the function's Hessian matrix, where $v_{max}$ and $v_{min}$ are the two eigenvalues. Besides the aspect ratio, the optimal approximations require different size allocation schemes on these linear elements for different $p$ values in the $L^{p}$ metric. It is shown the $L^{p}$ error measurements on these optimal linear elements should be equidistributed~\cite{Chen:07}.

For arbitrary surfaces, Amenta and Bern~\cite{Amenta98} pointed out the Delaunay triangulations of the $\epsilon$-samplings from the surface are good approximations and always have the correct topologies. In practice, due to the fact that $\epsilon$-sampling usually requires an excessive sampling of the surface, the anisotropic metric based methods and the proxy-driven methods become the two mainstream approaches to handle arbitrary surface approximation.

\subsection{Anisotropic Metric Based Methods}
\label{sec::related_works_anisotropic}
Centroidal Voronoi Tessellation (CVT) is widely used for nearly-equilateral mesh generation. It is a special type of Voronoi Diagram (VD)~\cite{Leibon:2000:DTV} whose sites coincide with the centroids of Voronoi cells. When a Riemannian metric tensor field is endowed to measure the way distance and angle are distorted, CVT is generalized to Anisotropic CVT (ACVT)~\cite{Du:2005:ACVT} to take the anisotropy into account. 

ACVT offers a flexible way to control the alignment, shape and size of the mesh elements. Several anisotropic metric based remeshing methods are proposed to avoid the time-consuming computation of the Anisotropic Voronoi Diagram (AVD)~\cite{Du:2005:ACVT}. These methods can be coarsely grouped into four categories: Delaunay refinement methods~\cite{Dobrzynski:2008,Boissonnat:13,Boissonnat:2008}, anisotropic CVT methods~\cite{Valette:2008:GRT,levy:12}, particle-based methods~\cite{SHIMADA00, Zhong:2013:PAS}, Optimal Delaunay Triangulation (ODT) methods~\cite{Chen:07,Chen:04,Fu:14}. A more detailed comparison between these methods can be found in~\cite{Fu:14}.

When anisotropic metric based methods are used as the surface remeshing tool, the common approach~\cite{Valette:2008:GRT,Boissonnat:13,Fu:14} is to explicitly compute the curvature information, which is then used as the Hessian matrix to approximate a locally smooth function of the surface. It is worth noting that both the curvature estimation step~\cite{Valette:2008:GRT,Boissonnat:13} and the following tensor smoothing step~\cite{Boissonnat:13, Fu:14} may result in the error between the approximated smooth function and the surface. Thus even though anisotropic metric based methods usually generate high-quality meshes w.r.t. the input metric field, their surface approximation qualities can be impaired by this inaccuracy. In contrast, our approach is a proxy-driven method, which does not rely on the curvature tensor to specify the anisotropy. 

\subsection{Proxy-Driven Methods}
The proxy-driven approaches cast the problem of surface approximation to the proxy-based geometric partition. When geometric primitives with similar shape characteristics are grouped together, it forms a simplified representation of the original shape. The aspect ratio and region size of each cluster are expected to follow the requirement from the approximation theory. 

Given the surface $\mathcal{S}$ and the user-assigned partition number $n$, a partition is denoted as $\mathcal{P} = \lbrace \mathcal{C}_i \rbrace_{i=1}^n $, which satisfies $\mathcal{C}_i \cap \mathcal{C}_j = \emptyset$ for $i \neq j$, and $\cup_i\mathcal{C}_i = \mathcal{S}$. With a shape proxy~\cite{Cohen-Steiner:2004:VSA} associated with each cluster, proxy-driven methods define the optimal partition as the minimization of the proxy fitting energy function:
\begin{equation}
	\label{eqn:total_energy}
	E_{partition}( \mathcal{P} )  = \sum\limits_{i=1}^n E(\mathcal{C}_i),
\end{equation}
where $E(\mathcal{C}_i)$ denotes the proxy fitting energy of cluster $i$, i.e., usually the summation over the fitting errors from the shape proxy to all underlying primitives in $\mathcal{C}_i$.

For surface approximation, there are two widely used proxy-driven methods: the quadric error metric (QEM) method~\cite{Garland:1997:QEM, Heckbert99} and the plane fitting method~\cite{Garland:2001:HFC, Cohen-Steiner:2004:VSA}. QEM method~\cite{Garland:1997:QEM} uses a point $\mathbf{v}_i$ as the shape proxy for cluster $\mathcal{C}_i$. The geometric primitives for a cluster are tangent planes (points equipped with normal directions). For plane fitting methods, Dual QEM (DQEM)~\cite{Garland:2001:HFC} and VSA $L^2$ metric~\cite{Cohen-Steiner:2004:VSA} use a plane as the shape proxy. The fitting error is thus the integral over all the points to the corresponding plane proxy. Note both QEM and the plane fitting method are proved~\cite{Heckbert99, Cohen-Steiner:2004:VSA} to have the optimal asymptotic aspect ratio $\sqrt{\abs{k_{max}/ k_{min}}}$, where $k_{max}$ and $k_{min}$ are the two principal curvatures. On the other hand, their asymptotic cluster size behavior are not discussed in literature.

Besides the point proxy and the plane proxy, other non-planar proxies such as spheres, cylinders and rolling ball blend patches~\cite{Wu:2005:SR}, and quadric proxies \cite{Yan:06:QS} have been used for robust surface structure extraction. However, there is no evidence that their partitions have any relation to the approximation theory.
\section{Our Energy Definition}
\label{our_energy_definition}
\subsection{Background on Mahalanobis CVT}
\label{sec::relation_MCVT}
As Sec.~\ref{sec::related_works_anisotropic} shows, the estimated curvature tensor provides an inaccurate functional representation for the surface to be approximated. In contrast to traditional anisotropic metric based approaches, Richter and Alexa proposed the Mahalanobis CVT (MCVT)~\cite{Richter:2015:SMI}, where the anisotropic metric is learned from the embedding of the surface in ambient space.

In MCVT, the energy of each individual cluster is modeled by the integral of distances from an observation point to all the points w.r.t. a metric to be optimized:
\begin{equation*}
\underset{\mathbf{x}, \mathbf{M} = \mathbf{M}^{\top}, \abs{\mathbf{M}} = 1}{\text{min}} \iint\limits_{\mathcal{C}} (\mathbf{p}-\mathbf{x}) \mathbf{M} (\mathbf{p}-\mathbf{x})^{\top}   \;\mathrm{d}\mathbf{p},
\end{equation*}
where $\mathbf{x}$ and $\mathbf{M}$ are the observation point and the optimized metric correspondingly; $\mathbf{p}$ refers to the points in the cluster. Note the determinant of the optimized metric is constrained to be unity in MCVT.

It turns out this constraint directly leads to the observation points being the centroid $\mathbf{\bar{p}}$ and the optimal metric being the inverse covariance matrix normalized to have unit determinant:
\begin{equation*}
\label{eq:mcvt}
\begin{split}
\mathbf{x} &= \mathbf{\bar{p}} = \frac{\iint\limits_{\mathbf{p}\in\mathcal{C}} \mathbf{p} \mathrm{d}\mathbf{p}}{\iint\limits_{\mathbf{p}\in\mathcal{C}}\; \mathrm{d}\mathbf{p}}, \\
\mathbf{M}_{\text{MCVT}} &= \abs{\mathbf{U}(\mathcal{C})}^{\frac{1}{d}} \mathbf{U}^{-1}(\mathcal{C}),
\end{split}
\end{equation*}
where $\mathbf{U}(\mathcal{C})$ is the covariance matrix; $\abs{\mathbf{U}(\mathcal{C})}$ is its determinant and $d$ is the dimension of ambient space ensuring $\abs{\mathbf{M}_{\text{MCVT}}} = 1$.

\subsection{Ellipsoidal Variance Proxy}
\label{sec::optimal_aspect_ratio}
In this section, we propose an \emph{ellipsoidal variance proxy} as the learned geometric representation of the cluster. Denoted as ($\mathbf{\bar{p}}$, $\mathbf{U}(\mathcal{C})$), ellipsoidal variance proxy is used to measure the variability of $\mathcal{C}$ in ambient space:
\begin{equation*}
(\mathbf{x} - \mathbf{\bar{p}})^{\top} \mathbf{U}^{-1}(\mathcal{C}) (\mathbf{x} - \mathbf{\bar{p}}) = 1.
\end{equation*}
Denote the eigenvalues and eigenvectors of $\mathbf{U}(\mathcal{C})$ as $\mathbf{e}_j$ and $\lambda_j, j=1,2,3$. It can be seen this ellipsoid model depicts the variance of cluster points along the eigenvector directions: the ellipsoid's semi-principal axes along $\mathbf{e}_j$ has the length of $\sqrt{\lambda_j}$.

We focus the discussion on the squared volume of the ellipsoidal variance proxy. Since it provides a scalar measurement for the variance of $\mathcal{C}$ in the ambient space, we denote it as the \emph{overall variance}. It can be seen that the overall variance is mathematically equivalent (up to a constant factor) to the determinant of the covariance matrix $\abs{\mathbf{U}(\mathcal{C})}$:
\begin{equation*}
\label{eq:det_cov_matrix}
\begin{split}
\abs{\mathbf{U}(\mathcal{C})} &= \prod\limits_{j=1}^{3} \lambda_j = \frac{9}{16\pi^2} (\frac{4\pi}{3} \prod\limits_{j=1}^{3} \sqrt{\lambda_j})^2 \\
&= \frac{9}{16\pi^2} \times \text{squared ellipsoid volume}.
\end{split}
\end{equation*}

In the remainder of the section, we justify the overall variance is well suited for surface approximation from two aspects. Firstly, it has an ACVT nature to penalize a cluster falling into disconnected parts, i.e., this overall variance can be interpreted as measuring the distance of all the points from the centroid w.r.t. the metric $\mathbf{M}_{\text{PCA}}(\mathcal{C}) = \abs{\mathbf{U}(\mathcal{C})} \mathbf{U}^{-1}(\mathcal{C})$:
\begin{equation}
\label{eq:det_cov_acvt}
\begin{split}
&\iint\limits_{\mathcal{C}} (\mathbf{p}-\mathbf{\bar{p}}) ^{\top}
\mathbf{M}_{\text{PCA}}(\mathcal{C})
(\mathbf{p}-\mathbf{\bar{p}}) \; \mathrm{d}\mathbf{p} \\
=& \iint\limits_{\mathcal{C}} \Tr{[
	(\mathbf{p}-\mathbf{\bar{p}}) ^{\top}
	\mathbf{M}_{\text{PCA}}(\mathcal{C})
	(\mathbf{p}-\mathbf{\bar{p}})
	]}  \; \mathrm{d}\mathbf{p} \\
=& \iint\limits_{\mathcal{C}} \Tr\left[
	\mathbf{M}_{\text{PCA}}(\mathcal{C})
	(\mathbf{p}-\mathbf{\bar{p}})
	(\mathbf{p}-\mathbf{\bar{p}}) ^{\top}	
	\right]\;\mathrm{d}\mathbf{p} \\
=& \Tr\left[ \mathbf{M}_{\text{PCA}}(\mathcal{C}) \iint\limits_{\mathcal{C}}
	(\mathbf{p}-\mathbf{\bar{p}})
	(\mathbf{p}-\mathbf{\bar{p}}) ^{\top}	
	\;\mathrm{d}\mathbf{p} \right]  \\
=& \abs{\mathbf{U}(\mathcal{C})}.
\end{split}
\end{equation}
Similar to the MCVT\cite{Richter:2015:SMI}, its minimization will lead to locally connected patches asymptotically.

Secondly, in the asymptotic limit for twice-differentiable elliptic/hyperbolic surface regions, minimizing a cluster's overall variance results in the cluster's aspect ratio converges to the optimal aspect ratio for surface approximation. In the asymptotic limit, each cluster becomes an infinitesimal local patch centered around some point. Its covariance matrix is the integral over the local neighborhood. Thus the overall variance is a function of the surface curvature and local neighborhood shape. If we fix the neighborhood patch size and let it freely stretch along the principal directions, we show in Appendix~\ref{sec::asymptotically_convergence} that the minimization of the overall variance will drive the clusters' aspect ratio converging to $\sqrt{\abs{k_{max}/ k_{min}}}$, where $k_{max}$ and $k_{min}$ are the principal curvatures.

Analogous to the energy form of $\abs{\mathbf{U}(\mathcal{C})}$ as shown in Eq.~(\ref{eq:det_cov_acvt}), MCVT has the proxy-driven energy form of $\abs{\mathbf{U}(\mathcal{C})}^{\frac{1}{d}}$. Through the same framework of Appendix~\ref{sec::asymptotically_convergence}, it can be seen that MCVT also arrives at the optimal asymptotic aspect ratio. There is no need to introduce the unit determinant constraint by $d$ for surface approximation.

\subsection{PCA Energy Definition}
\label{sec::optimal_size}
Besides the asymptotic aspect ratio, the asymptotic element size also plays a crucial role for surface approximation. However, previous proxy-driven approaches provide no theoretic guarantee on their asymptotic cluster size behaviors. In this subsection, we propose a PCA-based energy with the optimal asymptotic size control. 

For surface approximation with respect to the $L^\infty$ metric, it is known that the global error equidistribution condition~\cite{Chen:07} asymptotically requires each cluster patch to have the orthogonal length proportional to $k_i^{-\frac{1}{2}}$ along its principal curvature direction, where the principal curvatures $k_i, i=1,2$ can vary on different surface regions (Sec.~2.2 and Sec.~3 of~\cite{Simpson:1994:AMT}).

To achieve the optimal asymptotic cluster size for $L^\infty$  metric, we define the \emph{PCA-based energy}, i.e., the covariance matrix's determinant normalized by the fourth power of the surface area:
\begin{equation}
\label{eqn::our_energy1}
E_{PCA}(\mathcal{C}) = \frac{\abs{\mathbf{U}(\mathcal{C})}}{\text{area}^4(\mathcal{C})}.
\end{equation}

To optimize the summation over all PCA energies, clusters compete with each other in the cluster size allocation scheme. For surface consisting of twice-differentiable elliptic and hyperbolic regions, we show in Appendix~\ref{sec::size_control} that the minimization of the summation over Eq.~(\ref{eqn::our_energy1}) will lead to the optimal asymptotic cluster size~\cite{Simpson:1994:AMT}.

\subsection{General Surface Handling}
\label{sec::degenerate_energy}
In this subsection, we discuss the PCA energy on general surface types, i.e., particularly on parabolic regions, planar regions and input surface as triangular meshes (densely sampled $\mathcal{C}^0$ surfaces).

\begin{figure}[!t]
	\centering
	\includegraphics[width=3.5in]{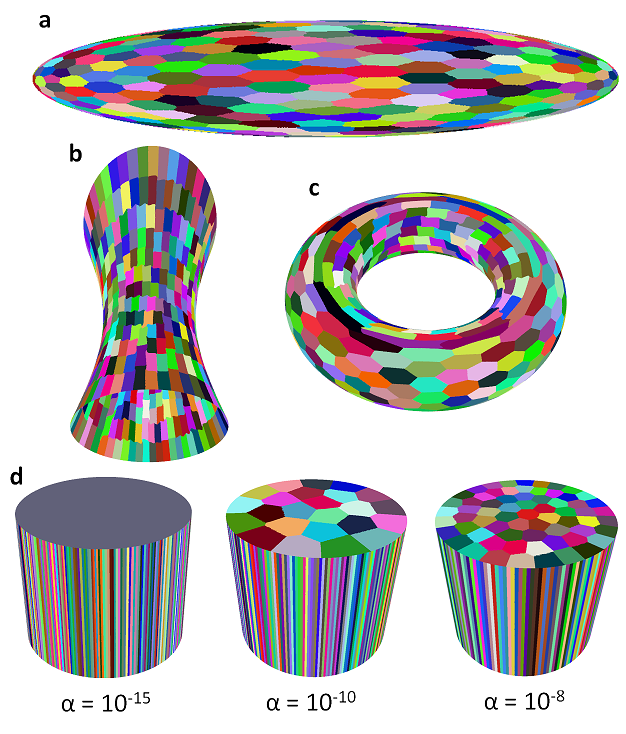}
	\caption{\label{fig:partition_simple_shapes} Partition result on simple models: (a) Partition (500 clusters) on the Ellipsoid model. (b) Partition (500 clusters) on the Hyperboloid  model. (c) Partition (500 clusters) on the Torus model. (d) Partition (300 clusters) on the Cylinder model using different $\alpha$ values.}
\end{figure}

\textbf{Parabolic Regions:} For parabolic regions, the aspect ratio of infinity is unreachable in practice. Meanwhile, this theoretic expectation can be seen as the limit of our asymptotic behaviors. In practice, the partition results are stretched along the degenerate direction of parabolic regions, as illustrated on the Cylinder model in Fig.~\ref{fig:partition_simple_shapes} (d).

\textbf{Planar Regions:} For planar regions, the determinant of cluster's covariance matrix is always zero, even for a cluster consisting of several disconnected patches. Thus, Eq.~(\ref{eqn::our_energy1}) may fail to produce a meaningful partition, which should be guaranteed even in the scenario of assigning a vertex/polygon budget on a pure planar surface. 

To overcome the possible degeneracies, MCVT~\cite{Richter:2015:SMI} redefines the covariance matrix to be a weighed average of the true covariance matrix and the identity matrix. For our consideration, we would like to keep the aspect ratio of other non-planar regions unchanged. To regularize the compactness of planar clusters, we adopt the CVT energy in a covariance matrix based form:
\begin{equation}
\label{eqn::our_energy2}
E_{PCA}(\mathcal{C}) = \alpha \cdot \Tr(\mathbf{U}(\mathcal{C})),
\end{equation}
where $\alpha$ is a small coefficient; $\Tr(\mathbf{U}(\mathcal{C}))$ is the CVT energy measured as the trace of the covariance matrix:
\begin{equation*}
\label{Eq::Spatial_Energy}
\begin{split}
&\quad \iint\limits_{\mathcal{C}} \norm{\mathbf{p} - \mathbf{\bar{p}}(\mathcal{C})}^2 d\mathbf{p} \\
&= \iint\limits_{\mathcal{C}} \Tr((\mathbf{p} - \mathbf{\bar{p}}(\mathcal{C})) (\mathbf{p} - \mathbf{\bar{p}}(\mathcal{C}))^{\top}) \;d\mathbf{p}\\
&= \Tr(\iint\limits_{\mathcal{C}} (\mathbf{p} - \mathbf{\bar{p}}(\mathcal{C}))^{\top} (\mathbf{p} - \mathbf{\bar{p}}(\mathcal{C}) \;d\mathbf{p}) \\
&= \Tr(\mathbf{U}(\mathcal{C})).
\end{split}
\end{equation*}

The coefficient $\alpha$ balances the relative cluster size between the planar regions and non-planar regions. Since the goal is to leave as much as possible vertex/polygon budget on non-planar regions, $\alpha$ is set as a very small constant $10^{-15}$. Generally speaking, for this choice, each planar region can be represented by one cluster when planar and non-planar regions are mixed (Fig.~\ref{fig:partition_simple_shapes} (d)). When a vertex/polygon budget is assigned on a pure planar surface, clusters can compete with each other and produce equilateral result (Fig.~\ref{fig:planar}).

To evaluate the cluster energy, we first check whether it is a planar region by testing whether $\abs{\mathbf{U}(\mathcal{C})} / \text{area}^5(\mathcal{C}) $ is smaller than $10^{-10}$. It is set as a small value to accept planar region with little tolerance. For planar regions, we use Eq.~(\ref{eqn::our_energy2}) to evaluate its energy with $\alpha=10^{-15}$. These parameters are fixed throughout all the experiments in this paper.

\textbf{Densely Sampled $\mathcal{C}^0$ Surfaces:} For practical purposes, the PCA-based energy needs to be evaluated on triangular meshes. For $\mathcal{C}^0$ surfaces, we noticed clusters are generally separated along sharp features (Fig.~\ref{fig:partition_feature_shapes}). After the introduction of our optimization algorithm, we provide a quantitative comparison in Sec.~\ref{sec::partition_results} on the asymptotic behavior with the theoretic expectation.

\section{Partition Optimization}
\label{sec::optimization_algorithm}
In this section, we develop a discrete merging-swapping variational framework to optimize the PCA-based energy. Our method does not rely on the concept of Voronoi cells and set the minimization of the PCA-based energy as the unique goal. In our discrete setting, clusters are built on individual triangular faces; and these triangular faces serve as the integral domain of the covariance matrix. Our partition algorithm consists of two steps: the merging step and the swapping step.

Initially, each face is treated as a cluster, which may form three cluster-pairs with its neighboring faces that share common edges. From the spirit of greedy pair merging~\cite{Garland:1997:QEM}, we mark cluster-pair merging $(\mathcal{C}_i, \mathcal{C}_j) \to \mathcal{C}_k$ with cost $E(\mathcal{C}_k) - E(\mathcal{C}_i) - E(\mathcal{C}_j)$. This cost can be interpreted as the increase of the total energy for this merging operation. We employ a min-heap to successively perform the least cost merging operation. Only a local update is needed to maintain the validity of the heap after each merging. 

Starting from the result of merging operations, the swapping step~\cite{Valette:2008:GRT} seeks the optimal partition by relaxing the face bindings from greedily merging. We let each triangle face freely decide which cluster to reside among the final $n$ clusters. For a triangle $\mathbf{T}$ residing in cluster $\mathcal{C}_i$, we can test whether swapping it to another cluster can decrease the energy in Eq.~(\ref{eqn:total_energy}). In practice, these swappings are launched in an iterative scheme. In each iteration, triangle $\mathbf{T}$ is limited to be the boundary triangles and only allowed to swap to their neighbors. The optimal partition is achieved when no swapping could further decrease the energy.

The optimization only involves with covariance matrix update and energy evaluation. As shown in Appendix \ref{sec::covariance_update}, only several float number multiplications are needed to update the covariance matrix by tracking the centroid and surface area of each cluster. Also, evaluating the determinant or the trace from a 3 $\times$ 3 matrix can be very efficient without the need to perform eigen-decomposition. Our merging-swapping framework shows much faster speed than the most widely used flooding framework~\cite{Cohen-Steiner:2004:VSA, Richter:2015:SMI}. The comparison and analysis can be found in Sec.~\ref{sec::partition}. 

\begin{figure}[!t]
	\centering
	\includegraphics[width=3.5in]{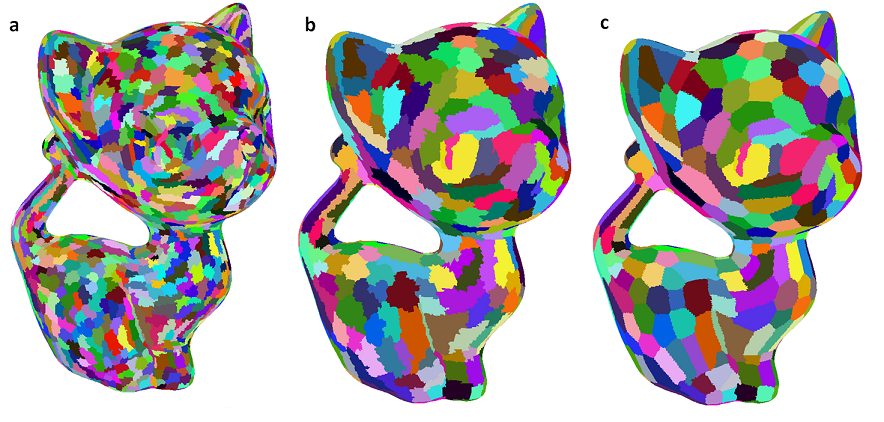}%
	\caption{\label{fig:illustrative_example} Illustrative example for our variational merging-swapping algorithm: (a) Greedily apply merging operation to 2,000 clusters. (b) Successively merge till 500 clusters. (c) Optimal partition for 500 clusters after applying swapping operations until convergence.}
\end{figure}

\begin{figure}[!t]
	\centering
	\includegraphics[width=3.5in]{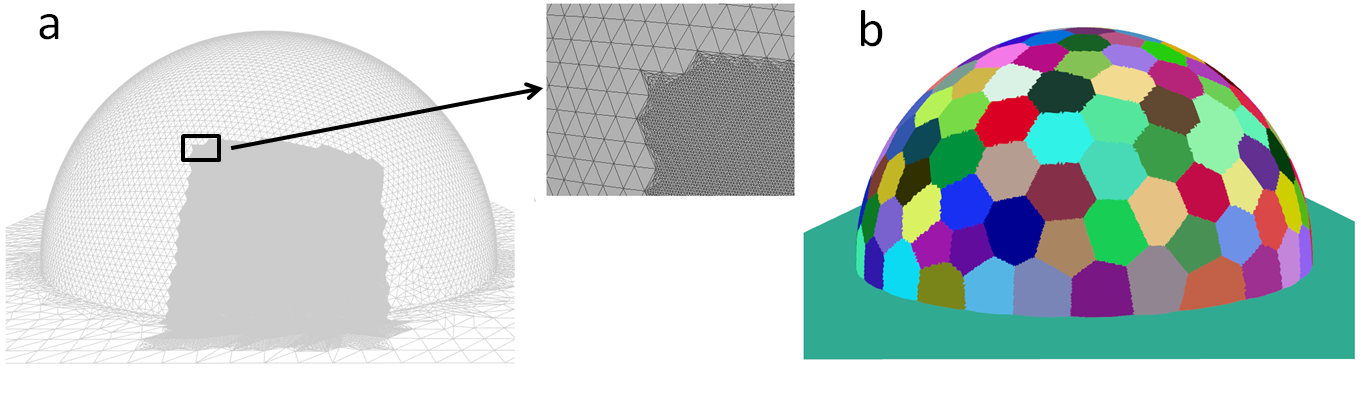}
	\caption{\label{fig:vary_density} Partition on model with different vertex density: (a) The half sphere on a plane with vary vertex density. (b) Despite the density change, the partition is uniform on the sphere and separated along the blended plane.}
\end{figure}

An illustrative example of our variational partition algorithm applied on the Kitten model is shown in Fig.~\ref{fig:illustrative_example}. Fig.~\ref{fig:vary_density} shows an example with varying sampling densities. Despite the 100 times density change, the partition is uniform on the sphere and separated along the blended plane. Our method is sampling-density-independent because the covariance matrix is the integral on the triangular patches.

\section{Partition on Triangular Meshes}
\label{sec::partition_results}
In this section, we evaluate the PCA-based energy on dense triangular meshes. These triangular meshes are densely sampled from the following twice-differentiable surfaces:
\begin{align*}
&\textbf{Ellipsoid: } &&\frac{x^2}{25}+y^2+z^2 = 1. \\
&\textbf{Hyperboloid: } &&\frac{x^2}{4}+y^2-z^2 = 1. \\
&\textbf{Torus: } &&(4 - \sqrt{x^2+y^2})^2 + z^2 = 4.
\end{align*}
Their $500$ cluster partition results are shown in Fig.~\ref{fig:partition_simple_shapes}(a), (b) and (c), respectively. We use the \emph{relative aspect ratio error} to measure the anisotropy deviation from the theoretic values:
\begin{equation*}
\delta r = \frac{r_m - r_t}{r_t},
\end{equation*}
where $r_m$ is the measured aspect ratio as the squared root of the ratio between the covariance matrix's first two eigenvalues, and $r_t$ is the theoretic aspect ratio at the projected cluster centroid.
\begin{figure}[!t]
	\centering
	\includegraphics[width=3in]{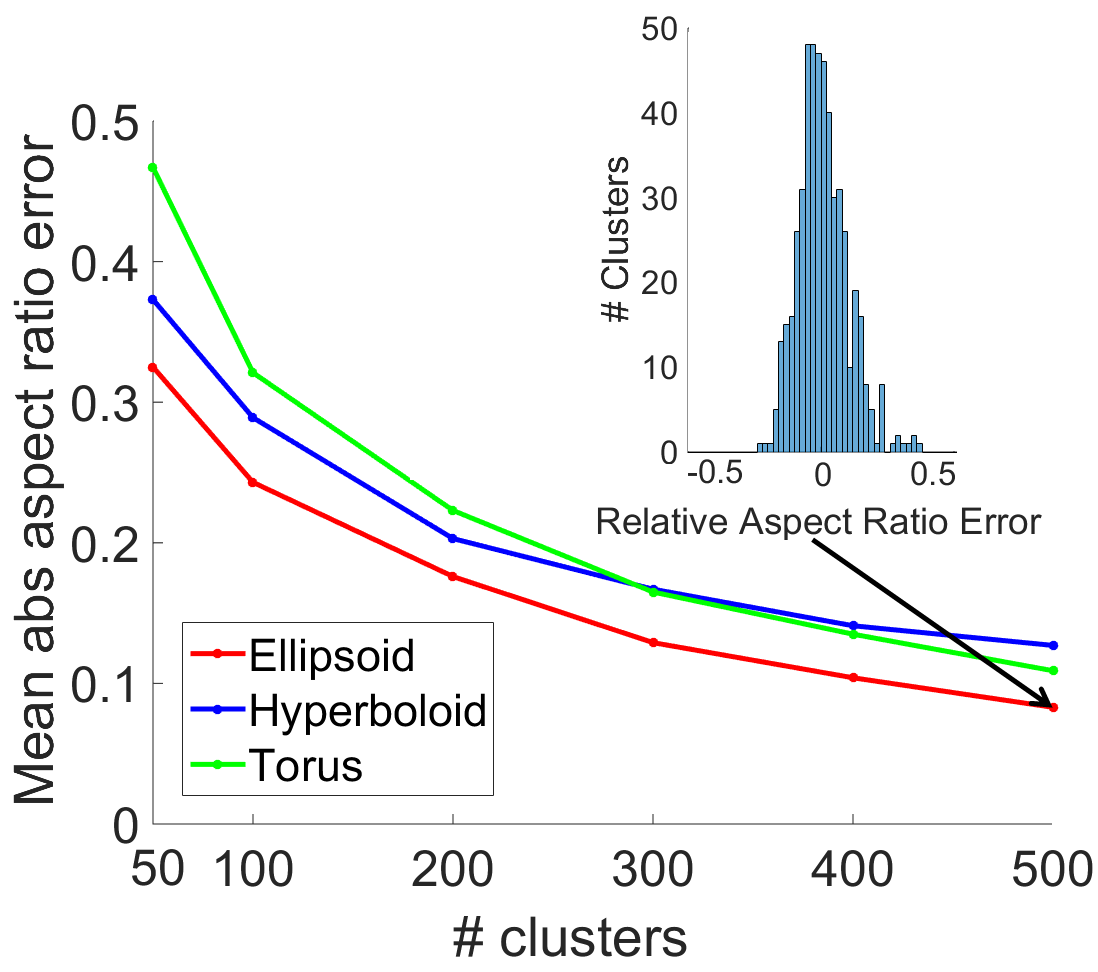}
	\caption{\label{fig:partition_convergence} The convergence curve w.r.t. the mean of the relative aspect ratio error's absolute value on the Ellipsoid mesh, Hyperboloid mesh, and Torus mesh.}
\end{figure}

Fig.~\ref{fig:partition_convergence} shows the convergence curve w.r.t. the mean of the relative aspect ratio error's absolute value. As the number of clusters reaches 500, the aspect ratio behavior is very close to the asymptotic theoretic expectation. Since the asymptotic cluster size is infinitesimal and not easy to compare directly, we illustrate this property through the global error equidistribution condition~\cite{Chen:07}. Fig.~\ref{fig:parabolid_cmp} and Fig.~\ref{fig:masque_cmp} show the nearly equidistributed result on the Paraboloid and Masque triangular mesh surface.

\begin{figure}[!t]
	\centering
	\includegraphics[width=3in]{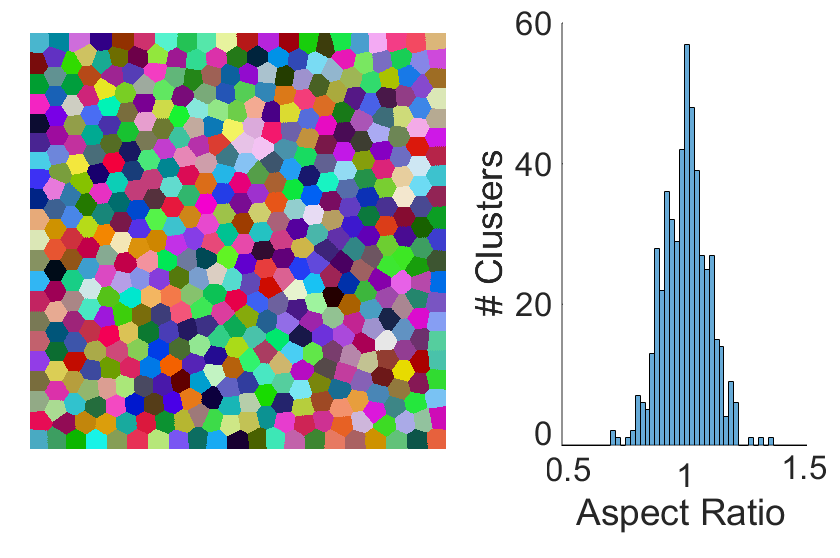}
	\caption{\label{fig:planar} Parition (500 clusters) on the Plane model, along with the aspect ratio histogram listed.}
\end{figure}

\begin{figure}[!t]
	\centering
	\includegraphics[width=3.5in]{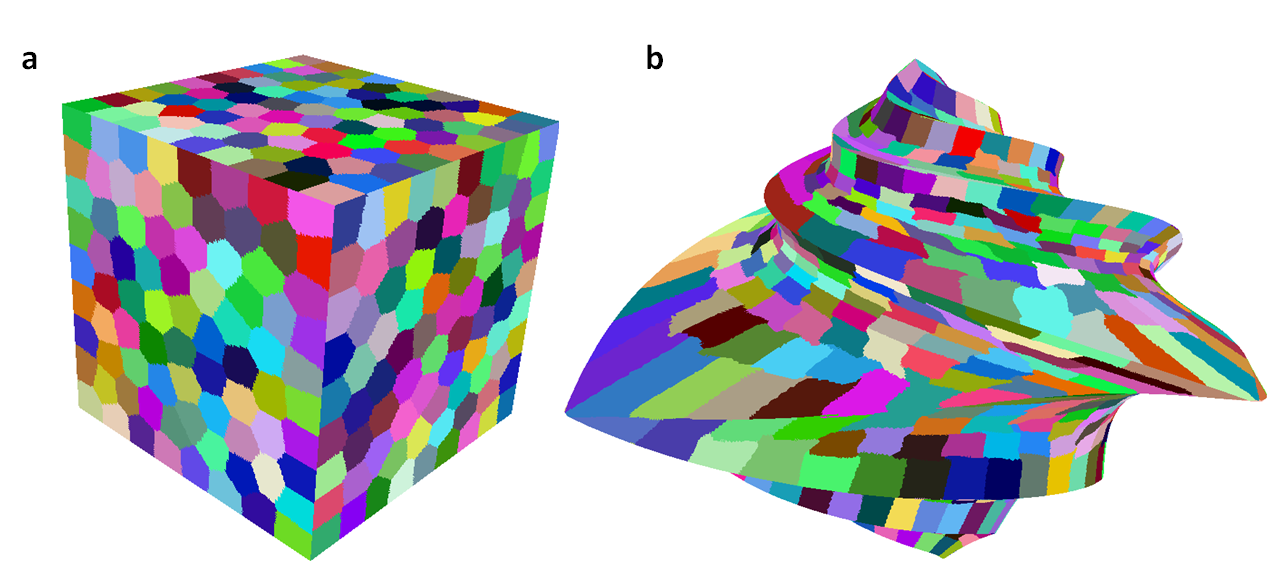}
	\hspace{1.5in}\parbox{3.5in}{\caption{\label{fig:partition_feature_shapes} Partition result on models with sharp features: (a) Partition (500 clusters) on the Cube model. (b) Partition (1000 clusters) on the Octa-flower model.}}
\end{figure}

For parabolic regions on the Cylinder model shown in Fig.~\ref{fig:partition_simple_shapes}(d), the clusters are elongated along the degenerate direction. Also we can see that $\alpha$ balances the relative size between the Cylinder sides and its planar ends. $\alpha =10^{-15}$ is enough to represent the planar region by one cluster. For pure planar surfaces, Fig.~\ref{fig:planar} shows the partition result on the Plane model with nearly hexagon tiling. For models with sharp edges, Fig.~\ref{fig:partition_feature_shapes} and Fig.~\ref{fig:fandisk} clearly show feature edges can be captured effectively by our PCA-based energy without any preprocessing. Generally, clusters are separated along the feature edges to avoid the sudden increment on the overall variance measure.

\section{Application to Remeshing}
\label{sec::our_meshing_algorithm}
\subsection{Post-Processing for Remeshing}
\begin{figure}[!t]
	\centering
	\includegraphics[width=3.0in]{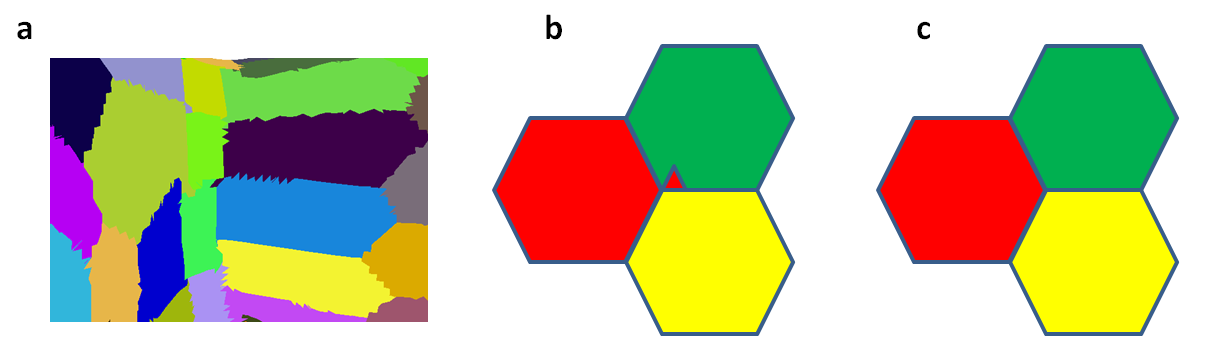}
	\caption{\label{fig:partition_cleaning} The post-processing on partition: (a) The disconnectedness after discrete optimization. (b) and (c) illustrate merging the red cluster's disconnected component to its least cost neighbor.}
\end{figure}

Disconnected clustering is penalized by the PCA-based energy. Due to the discrete optimization, disconnectedness is observed on the partition results. Since the swapping operation tests an entire triangle at each time, some triangles may be accepted with a very small energy decrease, while their neighbors are rejected. Disconnectedness usually occurs at the corner triangles "hanging" at the cluster's main component, as illustrated in Fig.~\ref{fig:partition_cleaning}(a).

In practice, usually dozens of boundary faces need to be cleaned on a model with several million faces. It would be computational expensive to guarantee the connectedness in the optimization algorithm. Instead, we run post-processing on these cases. The main component of the cluster is fixed while all the ``hanging components" are allowed to merge with their direct neighbors. Using the merging operation defined in Sec.~\ref{sec::optimization_algorithm}, these disconnected components are merged to least cost neighbor directly.

\subsection{Polygonal Remeshing}
\label{sec::poly_mesh}
\begin{figure}[!t]
	\centering
	\includegraphics[width=3in]{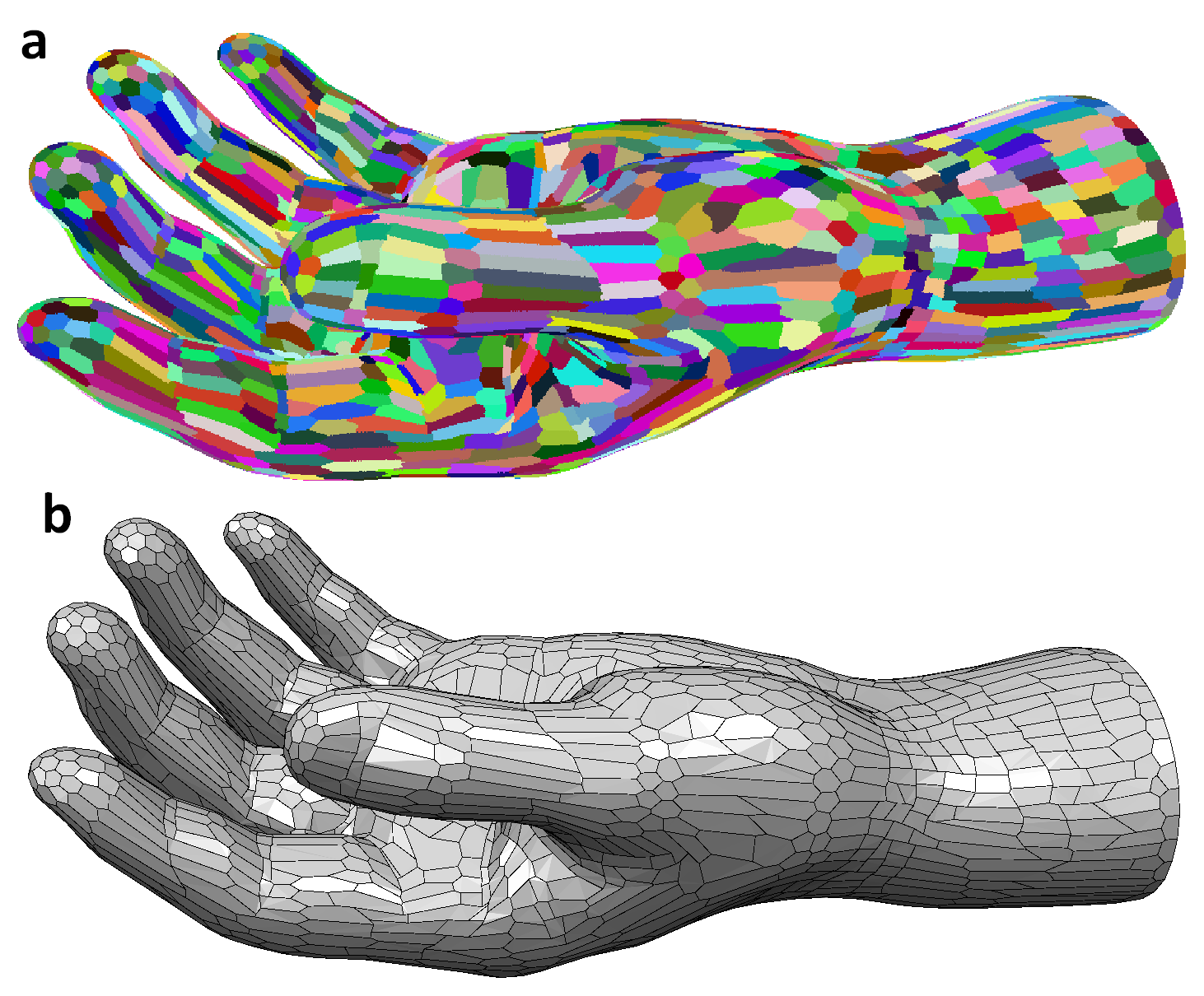}
	\caption{\label{fig:hand_partition} Polygonal mesh generation on the Hand model: (a) Partition the Hand model into 2,000 clusters. (b) Extracted 2,000 polygons from the optimal partition.}
\end{figure}

Due to its asymptotic behavior for piecewise-linear approximation, our partition result is straightforward to be extended to polygonal remeshing: First, each cluster is attached with a plane proxy defined by the centroid and the smallest eigenvector direction from the covariance matrix. Secondly, anchor vertices are created to represent the original vertices where three or more regions meet. For each anchor vertex, its spatial position is set as the average of all the projections from the original vertices onto its neighboring plane proxies. Thirdly, the anchor vertices are connected based on the neighboring cluster information and are sorted into a polygon with respect to the original mesh's orientation.

\begin{figure}[!t]
	\centering
	\includegraphics[width=3.3in]{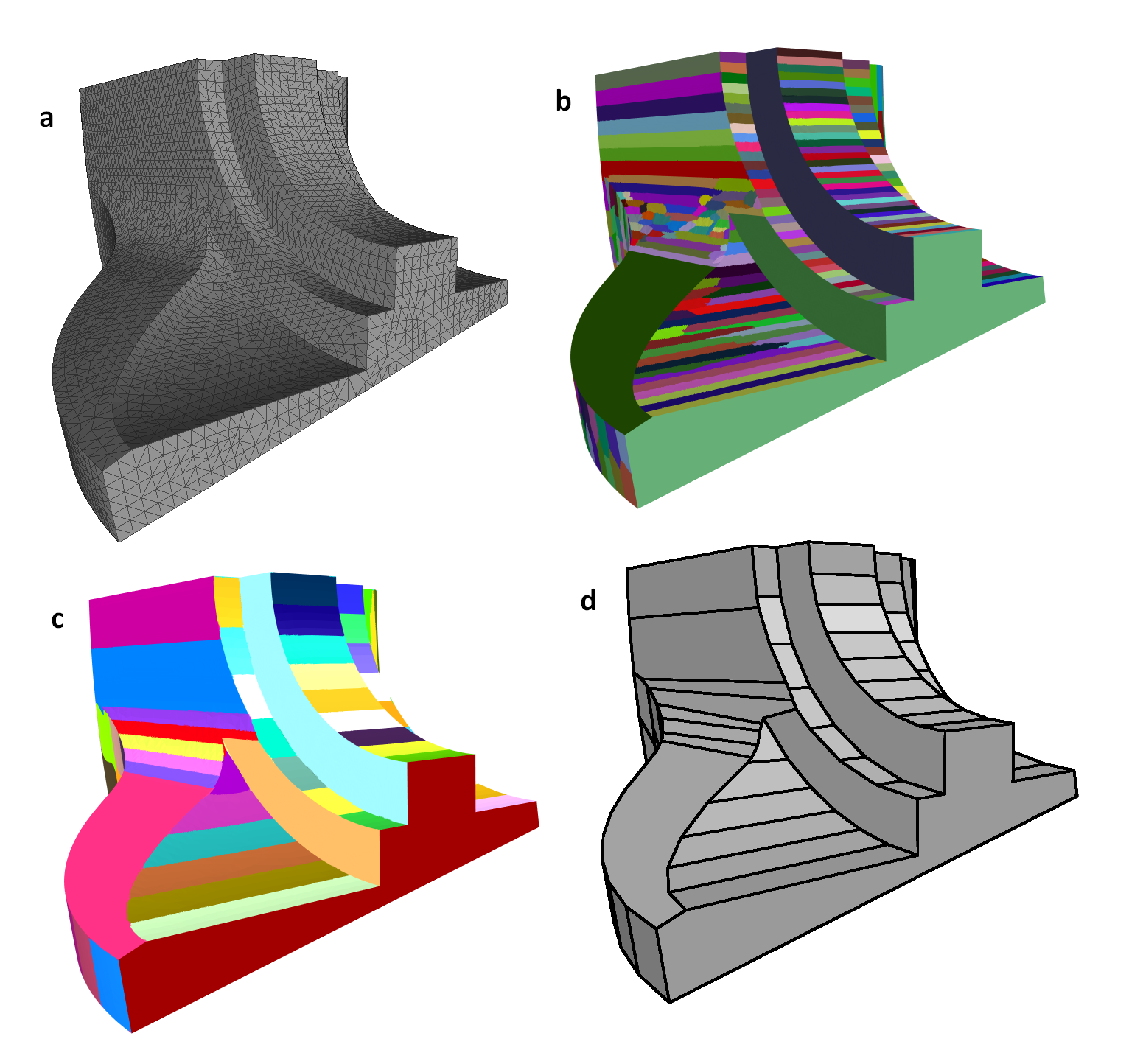}
	\caption{\label{fig:fandisk} Polygonal mesh generation on the Fandisk model: (a) The original Fandisk model. (b) 500 cluster partition result. (c) 100 cluster partition result. (d) Extracted 100 polygons. The feature lines are captured.}
\end{figure}

Fig.~\ref{fig:fandisk} illustrates the constructed polygons on the Fandisk model. On the original Fandisk model in Fig.~\ref{fig:fandisk}(a), the vertices of the blended regions are irregularly distributed. Our algorithm runs on the interpolated model (838k faces) using the midpoints of the edges. Note the blended regions contain non-neglectable local features due to the insufficient sampling of the initial model. When the cluster number is 500, since each patch is relatively small, the sampling pattern influences the partition result in Fig.~\ref{fig:fandisk}(b). When the cluster number becomes 100, the result is not significantly affected by these local features. For practical model with reasonable sampling rate, the cluster number is no longer limited to a small one.

Even though the constructed polygons are not necessarily flat, they are nearly flat as shown in Fig.~\ref{fig:hand_partition} and Fig.~\ref{fig:fandisk}. As quantitatively measured in planar hexagonal meshing~\cite{Li15}, the planarity error (PE) of each face is evaluated as the largest distance between any vertices and its fitting plane. In Tab.~\ref{tab:PE_statistics}, we provide the maximum and average planarity error measures in the unit of the diagonal of the model{\rq}s bounding box.

\begin{table}[!t]
	\centering
	\caption
	{Planarity error of the polygonal meshes. All the models are normalized to have a unit bounding box to measure errors.}
		\scalebox{0.8}{
			\begin{tabular}{lrrrrr}
				\toprule
				Model & Figure &\# Faces & \# Polygons & $PE_{avg}$ & $PE_{max}$ \\
				\midrule
				Hand  & Fig.~\ref{fig:hand_partition} & 4,403k & 2,000 & $7.15 \times 10^{-4}$ & $4.18 \times 10^{-3}$ \\ \midrule
				Ellipsoid  & Fig.~\ref{fig:tri_ellipsoid} & 327k & 500  & $1.27 \times 10^{-3}$  & $6.82 \times 10^{-3}$ \\ \midrule
				Fandisk & Fig.~\ref{fig:fandisk} & 838k & 100 & $3.21 \times 10^{-3}$ & $1.78 \times 10^{-2}$ \\ \midrule		
				Octa-Flower  & Fig.~\ref{fig:tri_octa_flower} & 995k & 1,000 & $1.10 \times 10^{-3}$ & $6.81 \times 10^{-3}$ \\ \midrule
				Paraboloid  & Fig.~\ref{fig:parabolid_cmp} & 524k & 500 & $1.66 \times 10^{-3}$ & $5.88 \times 10^{-3}$ \\ \midrule
				Masque  & Fig.~\ref{fig:masque_cmp} & 2,932k & 1,000 & $9.79 \times 10^{-4}$ & $4.52 \times 10^{-3}$ \\ \bottomrule
			\end{tabular}	
		}
	\label{tab:PE_statistics}
\end{table}

\subsection{Triangular Remeshing}
\label{sec::tri_mesh}
\begin{figure}[!t]
	\centering
	\includegraphics[width=3.0in]{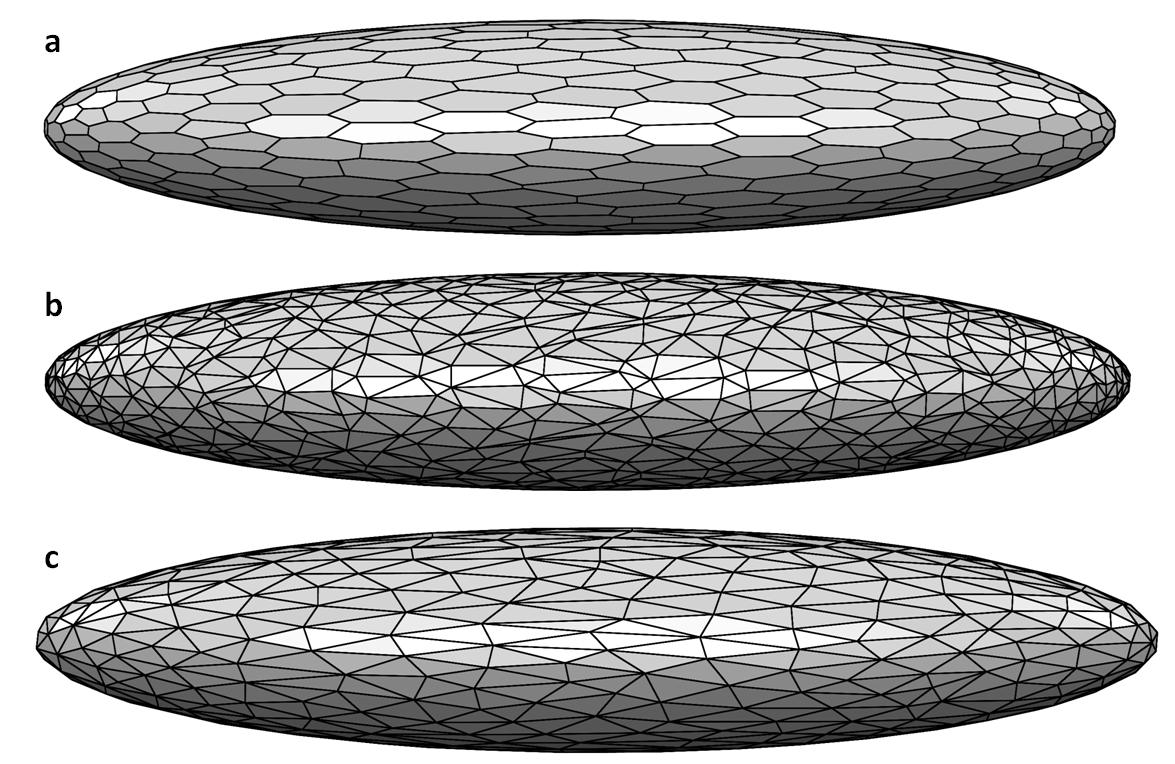}
	\caption{\label{fig:tri_ellipsoid} Triangular mesh generation on the Ellipsoid model: (a) 500 polygons. (b) The CDT mesh with 1,988 triangles. (c) The final mesh with 500 vertices after QEM simplification.}
\end{figure}
\begin{figure}[!t]
	\centering
	\includegraphics[width=3.5in]{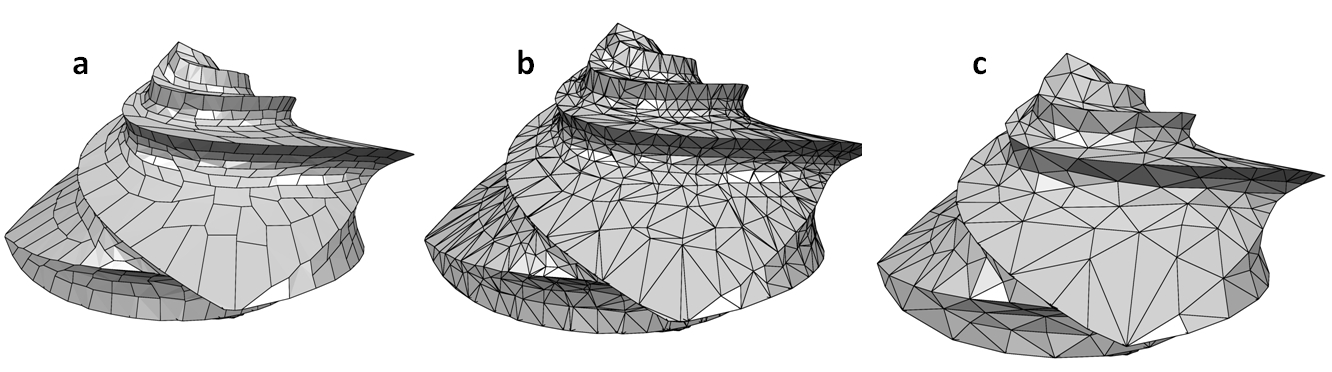}
	\caption{\label{fig:tri_octa_flower} Triangular mesh generation on the Octa-flower model: (a) 1,000 polygons. (b) The CDT mesh with 3,945 triangles. (c) The final mesh with 1,000 faces after QEM simplification. }
\end{figure}

Previous works~\cite{Valette:2008:GRT}~\cite{Du:1999:CVT} usually treat the partition result as the dual graph for triangular remeshing. However, utilizing a dualization step directly is not applicable for our partition result. Our partition cuts the input models along its feature edges (Fig.~\ref{fig:partition_feature_shapes}) and this dualization step based on the centroids will inevitably lose the control of the approximation at feature regions.

To generate a pure triangular representation, we first employ the “discrete” Constrained Delaunay Triangulation (CDT)~\cite{Cohen-Steiner:2004:VSA} to triangulate the polygons. It enforces the boundary edges of the existing polygons while triangulating the polygons in a Delaunay-like fashion. Then the technique of QEM~\cite{Garland:1997:QEM} is used to reach the assigned number of vertices/faces. 

We admit the generated triangular meshes inherit the suboptimal nature from QEM contraction~\cite{Garland:1997:QEM}. The sense of optimality in the partition and polygonal representation is no longer available in the triangular representation. On the other hand, the conversion is efficient as it starts from the simplified polygonal model. This process is illustrated on the Ellipsoid model (Fig.~\ref{fig:tri_ellipsoid}) and Octa-flower model (Fig.~\ref{fig:tri_octa_flower}).

\section{Experiments and Comparisons}
\label{sec::results}
In this section, we compare the surface approximation quality with the mainstream surface remeshing methods. Our algorithm is implemented using Microsoft Visual C++ 2010. The experiments are run on a desktop computer with Intel(R) Core i7-4770 CPU with 3.40GHz and 32GB DDR3 RAM.

The \emph{one-sided approximation error} from the Metro tool~\cite{Cignoni:98} is used for evaluation, i.e., for each vertex of the original mesh, its error distance on the generated simplified surface is computed. These error values are normalized w.r.t. the diagonal length of the original model's bounding box.

\subsection{Comparison on Proxy-Driven Partition}
\label{sec::partition}
In this subsection, we compare the partition result with QEM~\cite{Garland:1997:QEM} and VSA~\cite{Cohen-Steiner:2004:VSA}. Note two VSA energies are proposed in~\cite{Cohen-Steiner:2004:VSA}. VSA $L^{2}$ energy has the plane fitting form with the optimal asymptotic aspect ratio $\sqrt{\abs{k_{max}/ k_{min}}}$~\cite{Cohen-Steiner:2004:VSA}, while VSA $L^{2,1}$ energy has the imperfect aspect ratio $\abs{k_{max}/ k_{min}}$ for approximation. We focus our discussion on VSA $L^{2}$ energy in this subsection.

 \begin{figure}[!t]
 	\centering
 	\includegraphics[width=0.35\textwidth]{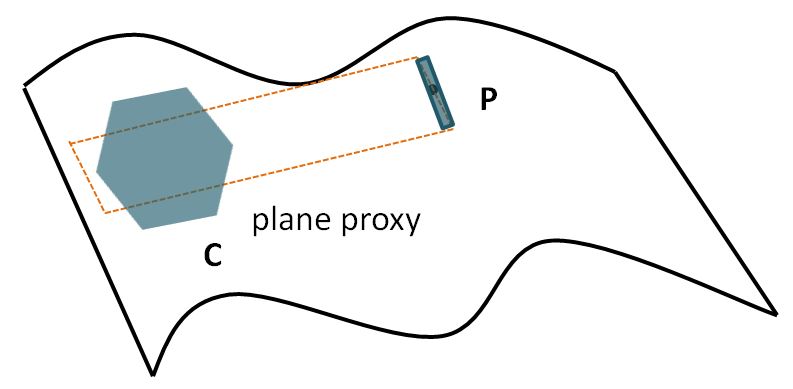}
 	\caption{\label{fig:planefitting_illustrate} Drawbacks of VSA $L^2$ energy: plane proxy will always "absorb" non-local points that it passes through.}
 \end{figure}

 \begin{figure}[!t]
 	\centering
 	\includegraphics[width=0.5\textwidth]{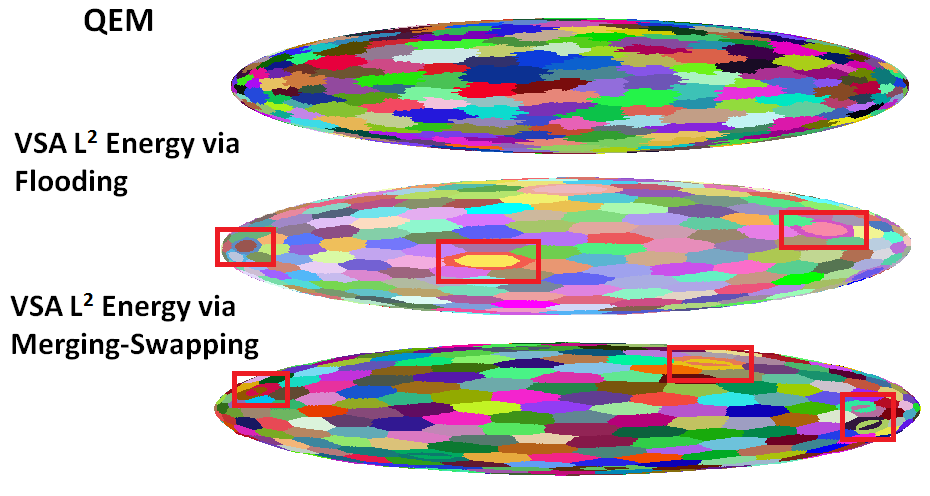}
 	\caption{\label{fig:others_result} Partition the Ellipsoid surface into 500 clusters using the QEM method~\cite{Garland:1997:QEM}, the VSA $L^2$ energy via flooding~\cite{Cohen-Steiner:2004:VSA} and the VSA $L^2$ energy via merging-swapping optimization. For the VSA $L^2$ energy, note that a cluster may totally reside in another for the flooding optimization, and may split into disconnected components in the merging-swapping scheme.}
 \end{figure}

As one of the proxy-driven energies, our PCA-based energy stands out due to its ACVT nature to constrain each cluster to be a locally connected patch. For QEM and VSA $L^{2}$ energy, locally connected patches are not the direct result from energy minimization. Instead, they require the optimization methods to ensure the connectedness of the clusters. As illustrated in Fig.~\ref{fig:planefitting_illustrate}, the minimization of the VSA $L^2$ energy will guide the clustering results to disconnected patches since a plane proxy will always "absorb" non-local points that it passes through. Likewise, the point proxy in QEM energy also absorbs non-local tangent planes. To ensure the cluster connectedness, QEM ends with suboptimal clustering through edge collapse~\cite{Garland:1997:QEM} and VSA $L^2$ energy uses a flooding optimization approach~\cite{Cohen-Steiner:2004:VSA}.

Our optimization process is an improvement over the flooding scheme: Firstly, the flooding scheme~\cite{Cohen-Steiner:2004:VSA} guarantees the cluster connectedness by
selecting the least cost flooding primitive from a priority queue. Denote \emph{N} as the number of primitives, the computation complexity for each iteration is \emph{O(NlogN)} since the selection from the priority queue costs \emph{O(logN)}. Meanwhile, our optimization framework is \emph{O(N)} with constant complexity for swapping tests. Secondly, the flooding algorithm suffers from the energy oscillations~\cite{Cohen-Steiner:2004:VSA} while we guarantee the monotonically decreasing, as only the swapping operations with energy decrease are accepted. As illustrated in Fig.~\ref{fig:masque_energy_cmp} and Tab.~\ref{tab:masque_statistics}, our framework requires much fewer iterations to converge.

Fig.~\ref{fig:others_result} shows the partition of the Ellipsoid surface by the QEM method, the VSA $L^2$ energy via flooding and the merging-swapping optimization scheme, respectively. Compared with Fig.~\ref{fig:partition_simple_shapes} (a), it clearly demonstrates the superiority of combining the PCA-based energy with the merging-swapping framework. QEM's greedy edge collapse scheme ends with suboptimal partition. For VSA $L^2$ energy, the connectedness via flooding scheme shows side effect: a cluster may totally resides in another cluster. On the other hand, our merging-swapping optimization, with a linear complexity, is not suitable for energies without disconnectedness penalty such as the VSA $L^2$ energy. The result in Fig.~\ref{fig:others_result} shows obvious disconnected components.

\subsection{Comparison on Polygonal Representation}
In this subsection, we compare the surface approximation quality on the polygonal mesh with MCVT~\cite{Richter:2015:SMI}, MCVT variant~\cite{Richter15} and VSA~\cite{Cohen-Steiner:2004:VSA}. The VSA $L^{2,1}$ results are used for remeshing purpose. The discrete CDT step~\cite{Cohen-Steiner:2004:VSA} is applied on all the polygonal meshes to generate the flat surfaces for approximation evaluation.

\begin{figure}[!t]
	\centering
	\includegraphics[width=3.5in]{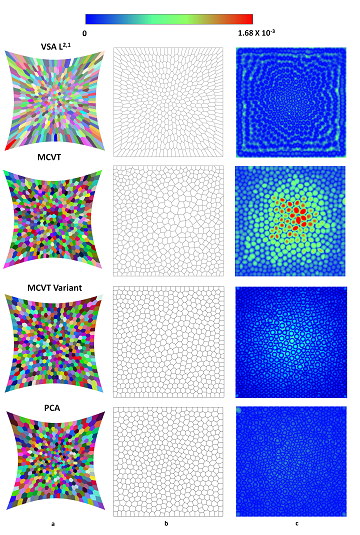}
	\caption{\label{fig:parabolid_cmp} Comparison of the polygonal meshes (500 polygons) on the paraboloid surface: (a) Partition result (top view). (b) The polygons mapped onto the $x\text{-}y$ domain. (c) The height difference between the interpolated height and the ground truth.}
\end{figure}

The first comparison is a dense sampling on a paraboloid function:
\[
z = x^2+y^2, \qquad -1 \leq x,y \leq 1.
\]
Since the Hessian of this function is constant everywhere, the optimal approximation requires each partition region has the same region size with the aspect ratio of $1$. Our partition result shows a nearly hexagonal tiling of the original surface in Fig.~\ref{fig:parabolid_cmp}. Since the VSA $L^{2,1}$ approach and MCVT provide no guarantee on optimal asymptotic cluster aspect ratio or size, higher approximation errors can be seen in over stretched regions of VSA and larger regions of MCVT. Meanwhile, MCVT variant~\cite{Richter15} scales the metric along the patch's normal direction. It provides some control over cluster size by homogenize the variance in the normal direction. Its result has the closest approximation quality to our method. The equidistributed height approximation error of our method is consistent with the theoretical proof in Appendix~\ref{sec::size_control}.

\begin{figure}[!t]
	\centering
	\includegraphics[width=3.2in]{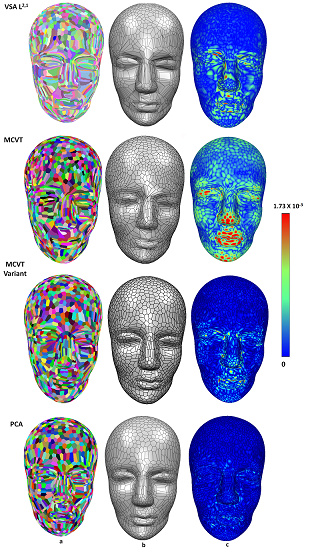}
	\caption{\label{fig:masque_cmp} Comparison of the polygonal meshes (1,000 polygons) on the Masque model: (a) Partition result. (b) The polygonal mesh. (c) Color-encoded one-sided approximation error distribution on the original mesh.}
\end{figure}

\begin{figure}[!t]
	\centering
	\includegraphics[width=2.8in]{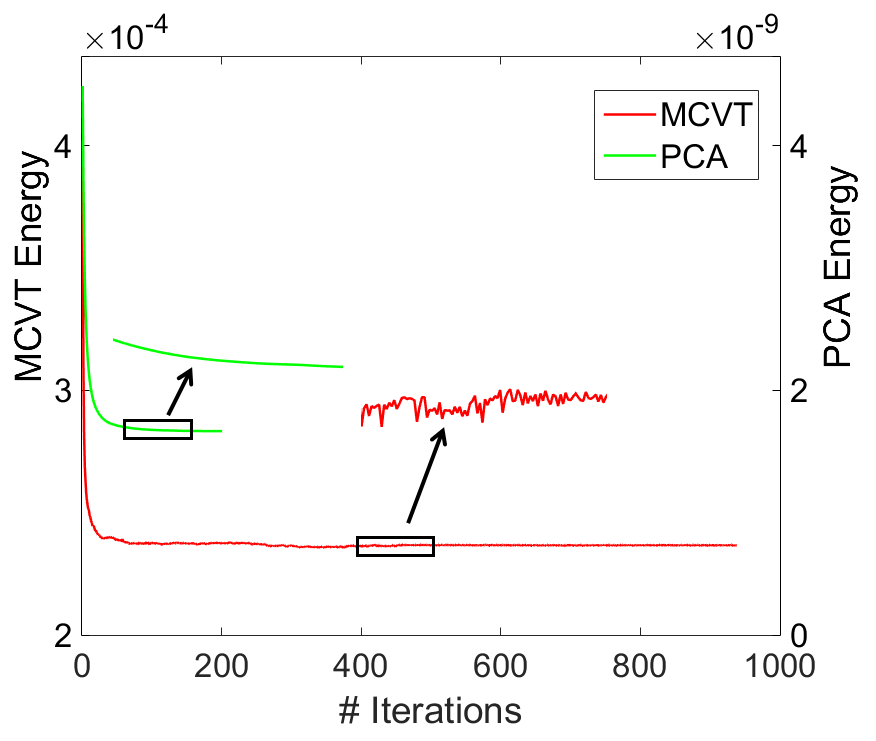}
	\caption{\label{fig:masque_energy_cmp} Comparison of the optimization convergence on the Masque model (1,000 polygons) between MCVT and PCA. The flooding scheme of MCVT suffers from energy oscillations, as shown in the zoom in views stretched 500 times along the Y axis. }
\end{figure}
\begin{table}[!t]
	\centering
	\caption
	{Polygonal mesh approximation quality measured after CDT triangulation.}
	\scalebox{0.8}{
		\begin{tabular}{lrrr}
			\toprule
			Model  & Mean dist. & RMS dist. & Hausdorff dist. \\
			& ($10^{-4}$) & ($10^{-4}$) & ($10^{-4}$)\\
			\midrule
			Paraboloid (VSA $L^{\scalebox{0.6}{2,1}}$)  & 1.36 & 1.74 & 27.45\\
			Paraboloid (MCVT) & 3.91 & 4.72 & 25.59 \\
			Paraboloid (MCVT Variant) & 1.16 & 1.46 & 8.27 \\
			Paraboloid (PCA)  & \textbf{0.93} & \textbf{1.26} & \textbf{5.38}\\
			\midrule
			Masque (VSA $L^{\scalebox{0.6}{2,1}}$)  & 1.81 & 2.83 & 24.28\\
			Masque (MCVT)  & 2.97 & 4.25 & 42.43 \\
			Masque (MCVT Variant) & 1.28 & 1.88 & 28.74 \\
			Masque (PCA) & \textbf{0.72} & \textbf{1.23} & \textbf{8.87}\\
			\bottomrule
		\end{tabular}
	}		
	\label{tab:poly_statistics}
\end{table}
\begin{table}[!htb]
	\centering
	\caption
	{Time statistics for the polygonal mesh generation.}
	\scalebox{0.8}{
		\begin{tabular}{lrrrr}
			\toprule
			Model (\# Vert.)  & \# Partition &  \multicolumn{3}{c}{Time(s)/ \# Iter} \\ 
			\cline{3-5} \\ 
				   &              & VSA & MCVT & PCA\\
			\midrule
			Ellipsoid (164k)  & 500 & 212.19/643  & 117.3/846  & \textbf{11.0/107} \\ \midrule
			Paraboloid (263k) & 500 & 296.2/479 & 255.7/662 & \textbf{27.8/169} \\ \midrule
			Masque (1,467k)  & 200 & 5,197.5/1,871 & 681.4/427 & \textbf{203.5/213} \\ \midrule
			Masque (1,467k)  & 1,000 & 5,984.3/2,000 & 1632.5/939 & \textbf{190.93/198} \\ \midrule
			Masque (1,467k)  & 5,000 & 6,578.5/2,000 & 2,246.8/1,687 & \textbf{173.67/166} \\ \bottomrule
		\end{tabular}
	}		
	\label{tab:masque_statistics}
\end{table}

Fig.~\ref{fig:masque_cmp} shows the comparison on the Masque surface with the one-sided approximation errors. Besides the more consistent polygonal anisotropy, our method shows better approximation quality at regions where the anisotropy changes rapidly. The nearly equidistributed approximation errors at the regions around the eyes, nose and lips clearly demonstrate the power of the PCA-based energy. Tab.~\ref{tab:poly_statistics} lists the statistics of the mean distance, root mean square (RMS) distance and the Hausdorff distance.

Tab.~\ref{tab:masque_statistics} lists the computational time of polygonal mesh generation by these methods. We denote the ratio of energy change as $\frac{\abs{E^{i}-E^{i-1}}}{E^{i}}$, where $E^{i}$ is the energy sum over all clusters at i-\emph{th} iteration. The optimization algorithm stops when the ratio of energy change smaller than $10^{-5}$ or the iteration number reaches 2,000. Since our algorithm is monotonically decreasing, it requires fewer iterations to converge, as shown in Fig.~\ref{fig:masque_energy_cmp}. Generally speaking, our method is about 10 times faster than MCVT and about 20 times faster than VSA. Different with VSA and MCVT, as the number of partition increases, our algorithm converges even faster with fewer iterations.

\subsection{Comparison on Triangular Representation}
To evaluate the approximation quality of the pure triangular representation, we compare with the mainstream anisotropic remeshing methods. The anisotropic metric is used as the original papers recommended: The particle-based method~\cite{Zhong:2013:PAS} uses the 6D metric in terms of vertices position and normals while ACVT~\cite{Valette:2008:GRT} and LCT~\cite{Fu:14} estimate curvature tensors on the triangular meshes~\cite{Cazals:2005}~\cite{Rusinkiewicz:2004}~\cite{Boissonnat:13}.

\begin{figure}[!t]
	\centering
	\includegraphics[width=3.5in]{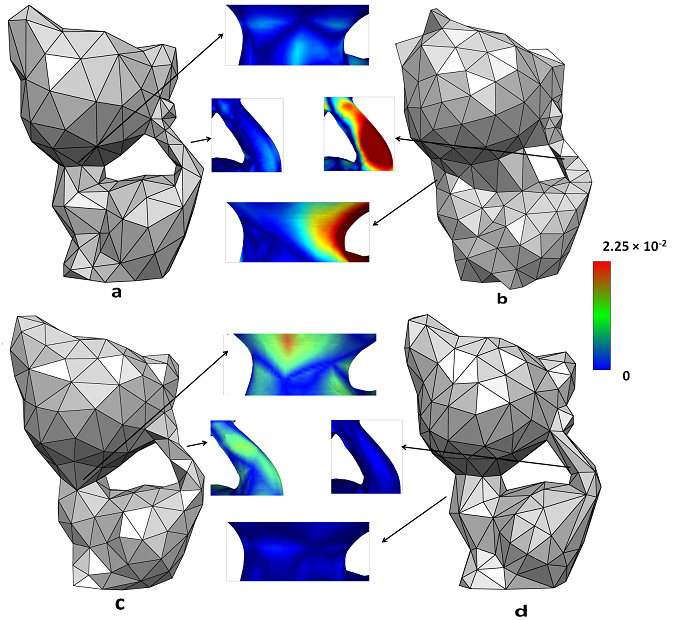}
	\caption{\label{fig:kitten_cmp} Comparison on the Kitten model with 200 vertex budget: (a) The ACVT approach. (b) The LCT approach. (c) The Particle-based approach. (d) Our result. Note the color-encoded error distribution is shown on the original model.}
\end{figure}

The comparison is performed on the Kitten model and the Buddha model. It should be noted the features of the models are not pre-tagged for the comparison, thus LCT and the particle-based method may miss the features during the projection step from the optimized vertices onto the triangular mesh, resulting in a much higher error than usual. The ACVT approach uses the QEM vertex placement scheme~\cite{Garland:1997:QEM} and handles the features well.

Fig.~\ref{fig:kitten_cmp} shows the approximation result on the Kitten model with 200 vertex budget. Our method has a powerful control over the feature regions. This can be seen from the alignment of the vertices over the neck and tail region. Obvious feature blurring can be observed in LCT and the particle-based method. Also our method provides the smallest approximation error as shown in the color-encoded error distribution on the original model.

\begin{figure}[!t]
	\centering
	\includegraphics[width=3.3in]{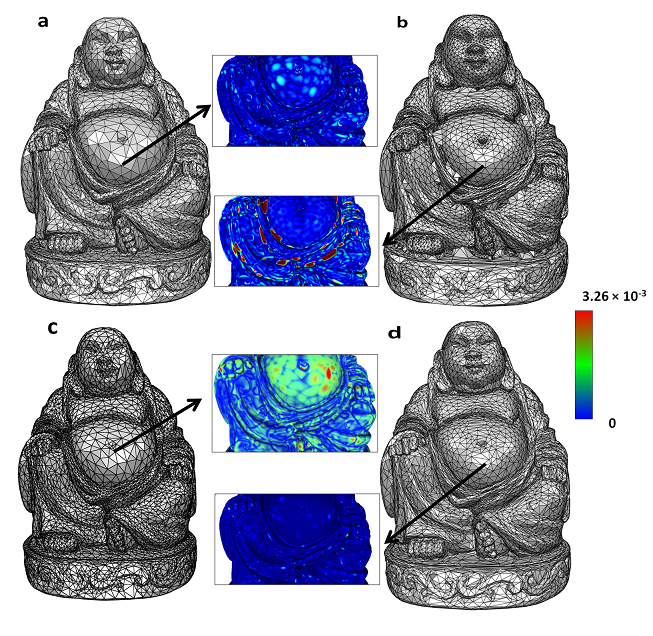}
	\caption{\label{fig:buddha_cmp} Comparison on the Buddha model with 8,000 vertex budget: (a) The ACVT approach. (b) The LCT approach. (c) The Particle-based approach. (d) Our result. Note the color-encoded error distribution is shown on the original model.}
\end{figure}

Fig.~\ref{fig:buddha_cmp} illustrates the approximation result on the Buddha model with $8,000$ vertex. Among all these methods, our PCA-based energy preserves the features on the base and the skirt, meanwhile exhibits the anisotropy with the smallest approximation error.

Compared with the anisotropic remeshing methods, our method provides the smallest one-sided approximation errors. In Tab.~\ref{tab:tri_statistics}, we can see that compared with ACVT, LCT and particle-based approach, our results shows improvement of the Hausdorff distance, one-sided mean distance and RMS distance. Our approximation quality is similar to the QEM result of MeshLab~\cite{Meshlab}. This is because applying QEM contraction on our polygonal mesh will inherit its suboptimal nature. Actually, it reveals the fact that applying QEM contractions on our polygonal representation has the same level of approximation quality as directly applying QEM on the original mesh.

\begin{table}[!t]
	\centering
	\caption
	{Triangular mesh approximation quality for different methods.}
	\scalebox{0.85}{
		\begin{tabular}{lrrr}
			\toprule
			Model  & Mean dist. & RMS dist. & Hausdorff dist. \\
			& ($10^{-3}$) & ($10^{-3}$) & ($10^{-3}$)\\
			\midrule
			Kitten (ACVT)  & 2.56 & 3.26 & 23.3\\
			Kitten (LCT) & 7.61 & 11.8 & 50.0 \\
			Kitten (Particle)  & 5.05 & 6.15 & 23.1\\
			Kitten (PCA) & \textbf{1.75} & \textbf{2.26} & 12.4\\
			Kitten (MeshLab) & 2.22 & 2.81 & \textbf{11.5} \\
			\midrule
			Buddha (ACVT)  & 0.27 & 0.38 & 3.26\\
			Buddha (LCT)  & 0.59 & 1.12 & 16.77\\
			Buddha (Particle) & 0.61 & 0.80 & 5.29\\
			Buddha (PCA) & \textbf{0.16} & \textbf{0.21} & \textbf{1.81}\\
			Buddha (MeshLab) & 0.17 & 0.22 & 2.06 \\
			\bottomrule
		\end{tabular}
	}		
	\label{tab:tri_statistics}
\end{table}

\subsection{Comparison on Noise Sensitivity}
\begin{figure}[!t]
	\centering
	\includegraphics[width=3.2in]{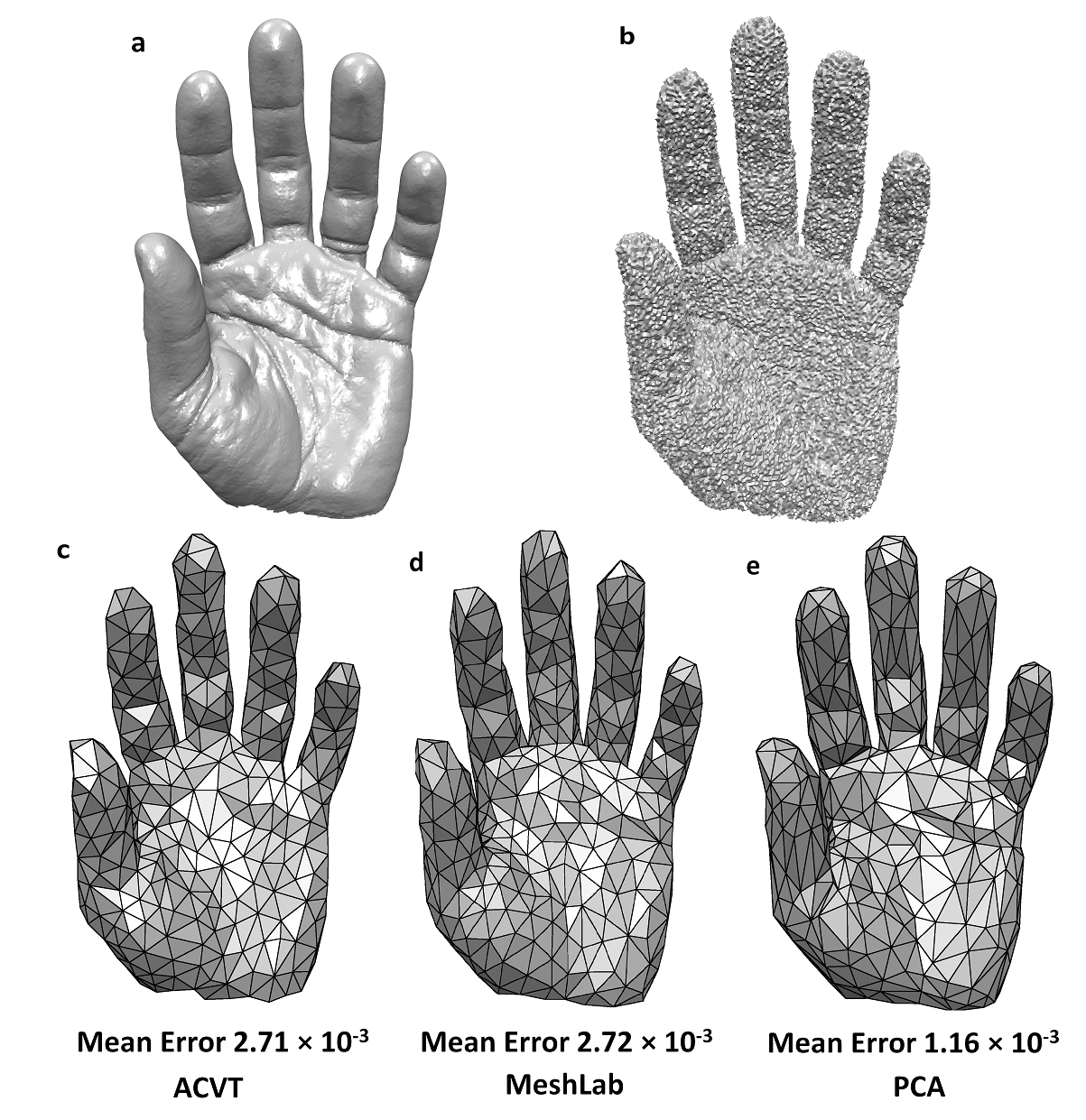}
	\caption{\label{fig:oliver_hand_noise} The Olivier hand surface: (a) The original model. (b) The noisy model. (c), (d) and (e) 500 vertices remeshing results on the noisy model using ACVT, MeshLab, PCA, respectively. The approximation error is measured on the original surface.}
\end{figure}
Since the model acquisition process of 3D scanners inevitably introduces noise signals, it would be useful to analysis the noise robustness of different methods. The surface approximation is directly performed on the noisy surface with widely used additive white Gaussian noise. The original surface is used as the ground truth for evaluation.

In Fig.~\ref{fig:oliver_hand_noise}, the noisy Olivier Hand model is generated by disturbing the vertices along their normal directions with white Gaussian noise whose variance is $0.002$ w.r.t. the diagonal length of the original model's bounding box. Fig.~\ref{fig:oliver_hand_noise}(c)(d)(e) show the remeshing results on the noisy model with $500$ vertices using ACVT, MeshLab and PCA, respectively. Among these results, our method provides the best visual result and still exhibits the anisotropy under severe additive noise. ACVT method requires the explicit curvature estimation, which is unreliable on noisy models. For the QEM energy used in MeshLab, the tangent planes of mesh vertices are also unstable under noise attack. On the other hand, instead of differentiating on the noisy surface, our PCA-based energy adopts the covariance matrix in the integral form, which shows the robustness against noise.

\begin{table}[!t]
	\centering\caption
	{Time statistics for triangular mesh generation.}
	\scalebox{0.8}{
		\begin{tabular}{lrrrrr}
			\toprule
			Model (\# Vert.) & \# Output & Merging & Swapping & Meshing & Total\\ 
			& vert. & time(s) & time(s) & time(s) & time(s) \\
			\midrule
			Olivier Hand (53k) & 500 & 0.90 & 1.25  & 0.56 & 2.71 \\ \midrule
			Kitten (137k) & 200 & 2.74 & 9.72 & 1.21 & 13.67\\ \midrule
			Octa-flower (498k) & 1,000 & 10.51 & 47.08 & 4.26 & 61.85\\ \midrule
			Buddha (1,224k) & 20,000 & 28.95 & 33.06 & 15.69 & 77.70 \\ \bottomrule	
		\end{tabular}		
	}
	\label{tab:pca_time}
\end{table}

\begin{figure}[!t]
	\centering
	\includegraphics[width=3.5in]{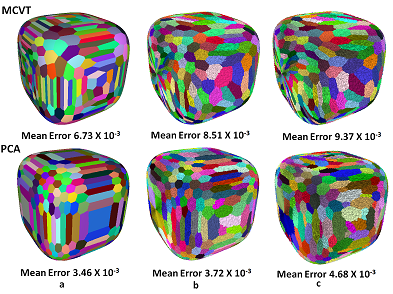}
	\caption{\label{fig:roundcube_noise} Partition results on the Rounded Cube model with Gaussian white noises whose variances are 0, 0.0015, 0.003, respectively from left to right. The one-sided mean errors are listed from the original model to their corresponding CDT meshes.}
\end{figure}

Fig.~\ref{fig:roundcube_noise} compares the partition stability with MCVT under different levels of noises. For smooth surfaces, MCVT gets comparable partition result with a different cluster size distribution. However, the learned metric is quickly governed by the noises. For our method, it starts from the initialization of the merging operations and keeps refining the partition by the swapping operations. Fig.~\ref{fig:roundcube_noise} (c) shows our method still exhibits robustness under white Gaussian noise with $0.003$ variance w.r.t. the bounding box's diagonal length.

\section{Conclusion and Discussion}
\label{sec::discussion}
In this paper, a surface partition approach is proposed to achieve the optimal asymptotic behavior in approximation theory. Different from previous proxy-driven approaches, our PCA-based energy has the ACVT nature to penalize the disconnectedness. It allows the more efficient merging-swapping optimization framework without the need to ensure the connectedness during optimization. 

The approximation quality is illustrated through the comparison of polygonal/triangular representations of different remeshing methods. When the application requires polygonal approximation or noise robustness approximation, our method outperforms others. Meshlab and our method are at the same level for triangular representation for non-noisy data.

Our partition result and polygonal approximation are guaranteed to provide the asymptotic optimal $L^\infty$ approximation. On the other hand, the triangular approximation lacks the optimality. The suboptimal nature inherited from the QEM contraction is a limitation of this work. Also, the QEM contraction does not detect or prevent global self-intersections of the generated meshes. The results can be benefited from extra computations, i.e., intermediate try-and-test steps. Starting from the optimal partition, improving the approximation quality for the triangular meshes will be the major direction for our future work.

Implicitly incorporating PCA over the surface patches is a powerful way to provide a concise geometric representation. Moreover, it seems to offer interesting alternatives for the approximation problems of other data types, such as the point cloud data, the images, etc. We will investigate these problems in the future.

\appendices
\section{Asymptotic Optimal Aspect Ratio}
\label{sec::asymptotically_convergence}
In this section, we prove that asymptotically on twice-differentiable elliptic/hyperbolic regions, minimizing the determinant of the covariance matrix will lead an infinitesimal patch $\mathcal{C}$ to have the optimal aspect ratio.

\begin{figure}[ht!]
	\centering
	\includegraphics[width=3.0in]{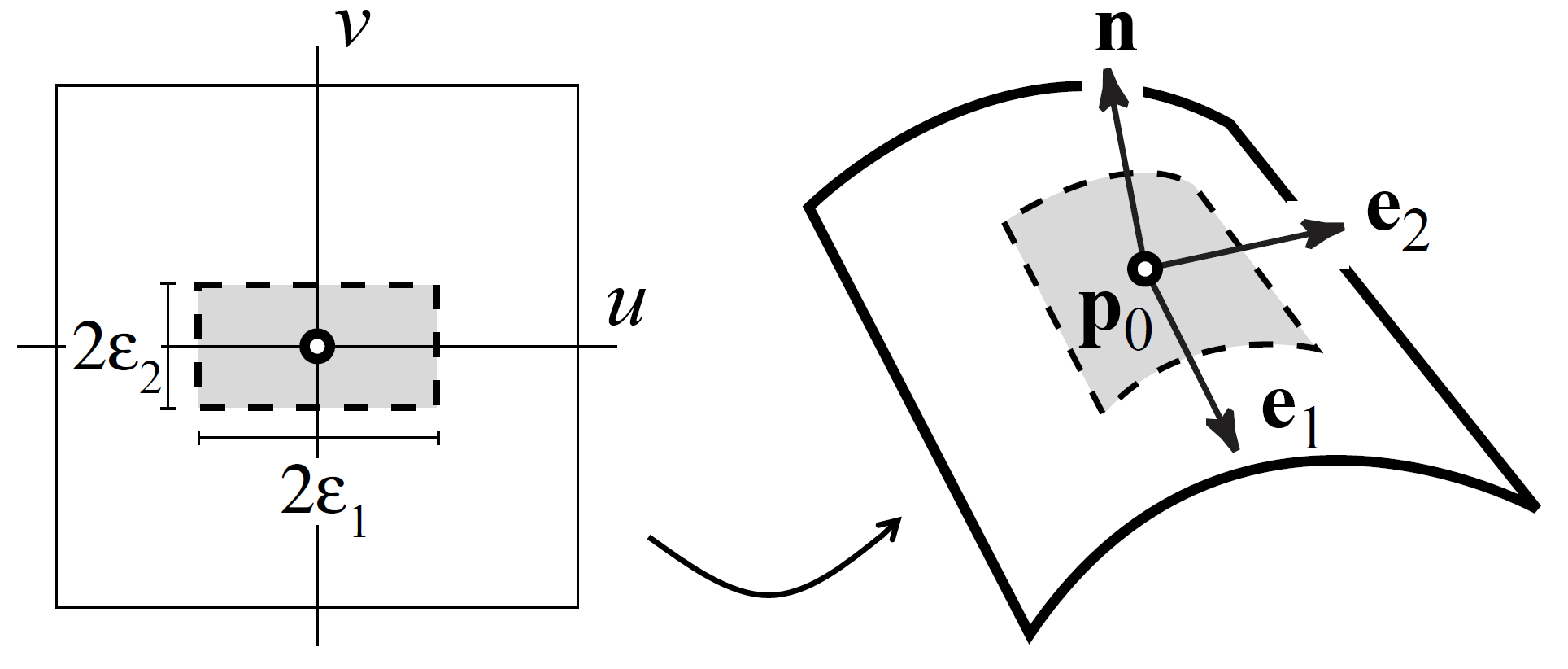}
	\caption{Local parameterization of point $\mathbf{p}_0$ on the surface, $\mathbf{u}$, $\mathbf{v}$ axes coincide with the principal axes $\mathbf{e}_1$ and $\mathbf{e}_2$.}
	\label{fig::aspect_ratio}
\end{figure}

Since the aspect ratio of a patch is involved, for the sake of simplicity, we now assume the infinitesimal patch is a rectangle parameter domain $\mathcal{C}$: $-\epsilon_1 \le u \le \epsilon_1, -\epsilon_2 \le v \le \epsilon_2$ around point $\mathbf{p}_{0}$ (see Fig.~\ref{fig::aspect_ratio}). Domain with other shapes would yield the identical result~\cite{Heckbert99}. In Fig.~\ref{fig::aspect_ratio}, we denote $\mathbf{e}_1$, $\mathbf{e}_2$ be the principal directions at $\mathbf{p}_0$. In the coordinate frame \lbrack$\mathbf{e}_1$, $\mathbf{e}_2$, $\mathbf{n}$\rbrack, the neighborhood of point $\mathbf{p}_0$ on the $\mathcal{C}^2$ surface can be approximated to the second order by:
\begin{equation}
\label{eq:p_approx}
\mathbf{p}(u,v) \approx \left(u, v, \frac{k_1u^2+k_2v^2}{2}\right)^{\top}.
\end{equation}
Denote the squared root of the rectangle area $\epsilon = \sqrt{\epsilon_1\epsilon_2}$, and the aspect ratio of the rectangle as $r = \epsilon_1/\epsilon_2$. The covariance matrix $\mathbf{U}(\mathcal{C})$ is the surface integral over the rectangular domain, which only depends on the unknowns $\epsilon$ and $r$:
\begin{equation*}
\mathbf{U}(\mathcal{C}) = \mathbf{U}(\epsilon, r) = \iint\limits_{\mathbf{p}\in\mathcal{C}} (\mathbf{p}-\bar{\mathbf{x}})(\mathbf{p}-\bar{\mathbf{x}})^{\top}   \mathrm{d}\mathbf{p}.
\end{equation*}
We now consider the problem: as $\epsilon_1$ and $\epsilon_2$ approach zero, what is the optimal aspect ratio $r$ that minimizes $\abs{\mathbf{U}(\epsilon, r)}$?

\textbf{Theorem 1} With $\epsilon_1,\epsilon_2 \rightarrow 0$ in the neighborhood of a $\mathcal{C}^2$ elliptic/hyperbolic region, the aspect ratio of the rectangle will automatically converge to $\sqrt{\left|\frac{k_2}{k_1}\right|}$ to minimize the overall covariance:
\begin{equation*}
r^{\star} = \argmin_{r} \abs{\mathbf{U}(\epsilon, r)} = \sqrt{\left|\frac{k_2}{k_1}\right|}.
\end{equation*}
\textbf{Proof} 
We first consider the centroid $\bar{\mathbf{x}}$ with the second order approximation in Eq.~(\ref{eq:p_approx}):
\begin{equation*}
\bar{\mathbf{x}}
= \frac{ \iint\limits_{\mathbf{p}\in\mathcal{C}}
	\left( \begin{array}{c}
	u \\
	v \\
	\frac{1}{2}(k_1u^2+k_2v^2) \end{array} \right)   \mathrm{d}\mathbf{p}}
{\iint\limits_{\mathbf{p}\in\mathcal{C}} \mathrm{d}\mathbf{p}} \\
= \left( \begin{array}{c}
0 \\
0 \\
\bar{\mathbf{x}}_n \end{array} \right),
\end{equation*}
making the substitution $\mathrm{d}\mathbf{p} =  \sqrt{1+k_1^2u^2+k_2^2v^2}\mathrm{d}u \mathrm{d}v$ into the third component of $\bar{\mathbf{x}}$:
\begin{equation}
\label{eq:xn_approx}
\begin{split}
\bar{\mathbf{x}}_n
& =
\frac{\iint\limits_{\mathcal{C}}
	\frac
	{1}{2}(k_1u^2+k_2v^2) \sqrt{1+k_1^2u^2+k_2^2v^2}
	\;\mathrm{d}u \mathrm{d}v}
	{\iint\limits_{\mathcal{C}} \sqrt{1+k_1^2u^2+k_2^2v^2}
	\;\mathrm{d}u \mathrm{d}v}  \\
& = \frac{\iint\limits_{\mathcal{C}}
	\frac
	{1}{2}(k_1u^2+k_2v^2) (1+ \mathbf{O}(u^2) + \mathbf{O}(v^2))
	\;\mathrm{d}u \mathrm{d}v}
	{\iint\limits_{\mathcal{C}} 1+ \mathbf{O}(u^2) + \mathbf{O}(v^2)
	\;\mathrm{d}u \mathrm{d}v}  \\
& = \frac{ \frac{2}{3}(k_1\epsilon_1^2+k_2\epsilon_2^2) (\epsilon_1\epsilon_2 + \mathbf{O}(\epsilon_1^3\epsilon_2) + \mathbf{O}(\epsilon_1\epsilon_2^3))}
{\epsilon_1\epsilon_2 + \mathbf{O}(\epsilon_1^3\epsilon_2) + \mathbf{O}(\epsilon_1\epsilon_2^3)}   \\
& \approx \frac{1}{6}(k_1\epsilon_1^2+k_2\epsilon_2^2).
\end{split}
\end{equation}
Note Eq.~(\ref{eq:xn_approx}) is the approximation for finite neighborhood, and becomes the exact limit for $\epsilon_1,\epsilon_2 \rightarrow 0$ with L'H{\^o}pital's rule.

Then the covariance matrix is the integral based on Eq.~(\ref{eq:p_approx}) and Eq.~(\ref{eq:xn_approx}):
\begin{equation*}
\begin{split}
\mathbf{U}(\mathcal{C}) = \iint\limits_{\mathcal{C}}
\left( \begin{array}{c}
u \\
v \\
w \end{array} \right)    	
\left( \begin{array}{ccc}
u &
v &
w\end{array} \right)
\sqrt{1+k_1^2u^2+k_2^2v^2} 	
\mathrm{d}u \mathrm{d}v,
\end{split}
\end{equation*}
where $w=\frac{1}{2}(k_1u^2+k_2v^2)-\frac{1}{6}(k_1\epsilon_1^2+k_2\epsilon_2^2)$.
All the non-diagonal terms of the covariance matrix are 0, since they are odd functions of u or/and v, and the three diagonal terms $U_{11}$, $U_{22}$, and $U_{33}$ correspond to the three eigenvalues:

\begin{equation}
\label{eqn:lambda1}
\lambda_1 = U_{11}
= \iint\limits_{\mathbf{p}\in\mathcal{C}} u^2 (1+ \mathbf{O}(u^2) + \mathbf{O}(v^2)) \mathrm{d}u \mathrm{d}v
\approx \frac{4}{3}\epsilon_1^3\epsilon_2,
\end{equation}
\begin{equation}
\label{eqn:lambda2}
\lambda_2 = U_{22}
= \iint\limits_{\mathbf{p}\in\mathcal{C}} v^2 (1+ \mathbf{O}(u^2) + \mathbf{O}(v^2)) \mathrm{d}u \mathrm{d}v
\approx \frac{4}{3}\epsilon_1\epsilon_2^3,
\end{equation}
\begin{equation}
\label{eqn:lambda3}
\begin{split}
\lambda_3 = U_{33}
& = \iint\limits_{\mathbf{p}\in\mathcal{C}} w^2
(1+ \mathbf{O}(u^2) + \mathbf{O}(v^2)) \mathrm{d}u \mathrm{d}v \\
& \approx \frac{4}{45}(k_1^2\epsilon_1^5\epsilon_2 + k_2^2\epsilon_1\epsilon_2^5)
\end{split}
\end{equation}
Note in Eq.~(\ref{eqn:lambda1}) and Eq.~(\ref{eqn:lambda2}) terms of degree six or higher in $\epsilon_1$ and $\epsilon_2$ are dropped, and terms of degree eight or higher are dropped in Eq.~(\ref{eqn:lambda3}). When $\epsilon_1,\epsilon_2 \rightarrow 0$, the overall variance is:
\begin{equation}
\label{eqn::cluster_determinant}
\abs{\mathbf{U}(\epsilon, r)} = \prod\limits_{j=1}^{3}\lambda_{j} = \frac{64}{405}\epsilon^{14} (k_1^2r^2+k_2^2r^{-2}).
\end{equation}
The optimal aspect ratio $r^{\star}$ can be reached with:
\begin{equation*}
\frac{\partial \abs{\mathbf{U}(\epsilon, r)}}{\partial r} |_{r=r^{\star}}
= \frac{128}{405}\epsilon^{14} (k_1^2 r^{\star} - k_2^2 (r^{\star})^{-3})
= 0.
\end{equation*}
For any elliptic or hyperbolic region, $r^{\star}$ is exactly the optimal aspect ratio for piecewise-linear approximation:
\begin{equation*}
\label{eqn::r_optimal}
r^{\star} = \argmin_{r} \abs{\mathbf{U}(\epsilon, r)} = \sqrt{\left|\frac{k_2}{k_1}\right|}. \rlap{$\qquad \Box$}
\end{equation*}

\section{Asymptotic Optimal Behavior for $L^\infty$ Approximation}
\label{sec::size_control}
In this section, we discuss the asymptotic optimal behavior for the proposed PCA energy. The notations in Appendix \ref{sec::asymptotically_convergence} is used. There are several approximations to discuss the PCA-based partition energy in Eq.~(\ref{eqn::our_energy1}). The cluster's surface area can be approximated by the rectangle patch's area $4\epsilon^{2}$ to the second order:
\begin{equation}
\label{Eq:area}
\begin{split}
\iint\limits_{\mathbf{p}\in\mathcal{C}} \mathrm{d}\mathbf{p} &= \iint\limits_{\mathcal{C}} \sqrt{1+k_1^2\,u^2+k_2^2\,v^2}
\;\mathrm{d}u \mathrm{d}v \\
&= 4\epsilon^{2} + \mathbf{O}(\epsilon^{4}).
\end{split}
\end{equation}
By dropping terms of degree four or higher in surface area, the partition energy becomes:
\begin{equation}
\begin{aligned}
\label{partition_energy}
& E_{partition}( \{\mathcal{C}_i \}_{i=1}^{n} ) = \sum\limits_{i=1}^n \frac{\abs{\mathbf{U}(\mathcal{C}_i)}}{area^4(\mathcal{C}_i)}  \\
&\approx \sum\limits_{i=1}^n \frac{1}{1620}(\epsilon^i)^{6}  \left((k_1^i)^2(r^i)^2+(k_2^i)^2(r^i)^{-2}\right),
\end{aligned}
\end{equation}
where the superscript $i$ refer to the index of the cluster.

Also, the total surface area constraint is:
\begin{equation}
\label{partition_constraint}
\text{subject to} \qquad
\sum\limits_{i=1}^n \iint\limits_{\mathbf{p}\in\mathcal{C}_i} \mathrm{d}\mathbf{p}
\approx \sum\limits_{i=1}^n 4(\epsilon^i)^{2} = A.
\end{equation}

For surface consisting of twice-differentiable elliptic/hyperbolic regions, we now prove this constrained optimization problem in Eq.~(\ref{partition_energy}) and Eq.~(\ref{partition_constraint}) will asymptotically lead each cluster's patch to have the orthogonal length proportional to $k_i^{-\frac{1}{2}}, i=1,2$ along the corresponding principal curvature direction, as required by~\cite{Simpson:1994:AMT}.

\textbf{Theorem 2} In the asymptotic limit on surface consisting of twice-differentiable elliptic/hyperbolic regions, the optimization of Eq.~(\ref{partition_energy}) with the constraint of Eq.~(\ref{partition_constraint}) will lead each cluster to have the orthogonal length proportional to $k_i^{-\frac{1}{2}}$:
\begin{equation*}
{\epsilon_1^i}^{\star} \sim \frac{1}{\sqrt{k_1^i}},
{\epsilon_2^i}^{\star} \sim \frac{1}{\sqrt{k_2^i}}.
\end{equation*}
or equivalently:
\begin{equation*}
{r^i}^{\star} = \sqrt{\left|\frac{k_2^i}{k_1^i}\right|},
({\epsilon^i}^{\star})^2 \sim \frac{1}{\sqrt{k_1^i k_2^i}}.
\end{equation*}

\textbf{Proof} It is easy to see $\{r^i \}_{i=1}^{n}$ are unconstrained and they are independent of each other due to the fact:
\begin{equation*}
\frac{\partial (E_{partition}( \{\epsilon^i, r^i \}_{i=1}^{n} ))}{\partial r^i}
=
\frac{\partial (E_{PCA}(\epsilon^i, r^i))}{\partial r^i}.
\end{equation*}
Thus the optimal aspect ratio of each cluster is the same as illustrated in Theorem 1 in Appendix \ref{sec::asymptotically_convergence}:
\begin{equation*}
{r^i}^{\star} = \sqrt{\left|\frac{k_2^i}{k_1^i}\right|}.
\end{equation*}
The cluster energy of the $i$-th cluster becomes:
\begin{equation*}
E_{PCA}(\epsilon^i) = \frac{\abs{\mathbf{U}(\mathcal{C}_i)}}{area^4(\mathcal{C}_i)} = \frac{1}{810}k_1^i k_2^i (\epsilon^i)^{6}.
\end{equation*}
The constrained optimization problem only depends on the cluster size allocation scheme $\{\epsilon^i \}_{i=1}^{n}$:
\begin{equation*}
\begin{aligned}
& {\text{min}}
& & \sum\limits_{i=1}^n k_1^i k_2^i (\epsilon^i)^{6}, \\
& \text{subject to}
& & \sum\limits_{i=1}^n (\epsilon^i)^{2} = \frac{A}{4}.
\end{aligned}
\end{equation*}

We can use Lagrange multiplier to analyze the constrained minimization problem :
\begin{equation*}
f( \{\epsilon^i \}_{i=1}^n, \lambda) = \sum\limits_{i=1}^n k_1^i k_2^i (\epsilon^i)^{6} - \lambda \left(\sum\limits_{i=1}^n (\epsilon^i)^{2} - \frac{A}{4}\right).
\end{equation*}
For the optimal cluster size of $i$-th cluster, it must satisfy:
\begin{equation*}
\left.\frac{\partial f( \{\epsilon^i \}_{i=1}^n)}{\partial \epsilon^i} \right|_{\epsilon^i={\epsilon^i}^{\star}}  = 6 k_1^i k_2^i ({\epsilon^i}^{\star})^{5} - 2 \lambda {\epsilon^i}^{\star} = 0,
\end{equation*}
\begin{equation*}
\frac{\partial f( \{\epsilon^i \}_{i=1}^n)}{\partial \lambda}  = \sum\limits_{i=1}^n (\epsilon^i)^{2} - \frac{A}{4} = 0.
\end{equation*}
So $\lambda$ is a constant and the optimal cluster size is inversely proportional to the square root of Gaussian curvature:
\begin{equation*}
({\epsilon^i}^{\star})^2 = \sqrt{\frac{\lambda}{3k_1^i k_2^i}} \sim \frac{1}{\sqrt{k_1^i k_2^i}}. \rlap{$\qquad \Box$}
\end{equation*}

\section{Efficient Update of Covariance Matrix}
\label{sec::covariance_update}
Even though the centroid of a cluster changes during the merging and swapping operation, we show in this section the update of the covariance matrix can be as efficient as several float number multiplications, without the need to sum over all of the primitives in this cluster.

This efficient update requires keeping track of the surface area and the centroid of the clusters. We use $A(\mathcal{C}_i)$, $\bar{\mathbf{x}}(\mathcal{C}_i)$, $\mathbf{U}(\mathcal{C}_i)$ to denote the surface area, centroid and covariance matrix of $\mathcal{C}_i$ respectively. The efficient update operations are listed below. The complete derivation can be found in the supplementary material.

\subsection{Initialization of A Triangle}
For a triangle $\mathbf{T}$ with three vertices $\mathbf{v}_1$, $\mathbf{v}_2$, $\mathbf{v}_3$:
\begingroup\smallColSep
\begin{equation*}
\begin{split}
A(\mathbf{T}) &= \frac{1}{2}\norm{(\mathbf{v}_2- \mathbf{v}_1) \times (\mathbf{v}_3- \mathbf{v}_1)}. \\
\bar{\mathbf{x}}(\mathbf{T}) &= \frac{1}{3}(\mathbf{v}_1+\mathbf{v}_2+\mathbf{v}_3). \\
\mathbf{U}(\mathbf{T}) &= \frac{A}{36}
\left(\begin{array}{ccc}
\mathbf{v}_1 &
\mathbf{v}_2 &
\mathbf{v}_3
\end{array}\right)
\left( \begin{array}{rrr}
2 & -1 & -1 \\
-1 & 2 & -1 \\
-1 & -1 & 2 \end{array} \right)
\left(\begin{array}{ccc}
\mathbf{v}_1 &
\mathbf{v}_2 &
\mathbf{v}_3
\end{array}\right)^{\top}.
\end{split}
\end{equation*}
\endgroup

\subsection{Merging Operation}
For a merging operation $(\mathcal{C}_i, \mathcal{C}_j) \to \mathcal{C}_k$:
\begin{equation*}
\begin{split}
A(\mathcal{C}_k) &= A(\mathcal{C}_i) + A(\mathcal{C}_j). \\
\bar{\mathbf{x}}(\mathcal{C}_k) &= \frac{A(\mathcal{C}_i)\bar{\mathbf{x}}(\mathcal{C}_i) + A(\mathcal{C}_j)\bar{\mathbf{x}}(\mathcal{C}_j)}{A(\mathcal{C}_i) + A(\mathcal{C}_j)}. \\
\mathbf{U}(\mathcal{C}_k) &= \mathbf{U}(\mathcal{C}_i) + \mathbf{U}(\mathcal{C}_j) \\
&+A(\mathcal{C}_i)\,\left(\bar{\mathbf{x}}(\mathcal{C}_i)-\bar{\mathbf{x}}(\mathcal{C}_k)\right)\,\left(\bar{\mathbf{x}}(\mathcal{C}_i)- \bar{\mathbf{x}}(\mathcal{C}_k)\right)^{\top} \\
&+A(\mathcal{C}_j)\,\left(\bar{\mathbf{x}}(\mathcal{C}_j)-\bar{\mathbf{x}}(\mathcal{C}_k)\right)\,\left(\bar{\mathbf{x}}(\mathcal{C}_j)- \bar{\mathbf{x}}(\mathcal{C}_k)\right)^{\top}.
\end{split}
\end{equation*}

\subsection{Swapping Operation}
For a swapping operation which swaps a triangle face $\mathbf{T}$ from $\mathcal{C}_i$ to $\mathcal{C}_j$. Suppose after swapping, $\mathcal{C}_i$ becomes $\mathcal{C}_{i^{'}}$ and $\mathcal{C}_j$ becomes $\mathcal{C}_{j^{'}}$, i.e., $\mathcal{C}_i=\mathcal{C}_{i^{'}} \cup \mathbf{T}$, and $\mathcal{C}_{j^{'}}=\mathcal{C}_j \cup \mathbf{T}$:

\begin{equation*}
\begin{split}
A(\mathcal{C}_{i^{'}}) &= A(\mathcal{C}_i) - A(\mathbf{T}). \\
A(\mathcal{C}_{j^{'}}) &= A(\mathcal{C}_j) + A(\mathbf{T}). \\
\bar{\mathbf{x}}(\mathcal{C}_{i^{'}}) &= \frac{A(\mathcal{C}_i)\bar{\mathbf{x}}(\mathcal{C}_i) - A(\mathbf{T})\bar{\mathbf{x}}(\mathbf{T})}{A(\mathcal{C}_i) - A(\mathbf{T})}. \\
\bar{\mathbf{x}}(\mathcal{C}_{j^{'}}) &= \frac{A(\mathcal{C}_j)\bar{\mathbf{x}}(\mathcal{C}_j) + A(\mathbf{T})\bar{\mathbf{x}}(\mathbf{T})}{A(\mathcal{C}_i) + A(\mathbf{T})}. \\
\mathbf{U}(\mathcal{C}_{i^{'}}) &= \mathbf{U}(\mathcal{C}_i) - \mathbf{U}(\mathbf{T}) \\
&-A(\mathcal{C}_{i^{'}})(\bar{\mathbf{x}}(\mathcal{C}_{i^{'}})-\bar{\mathbf{x}}(\mathcal{C}_i))(\bar{\mathbf{x}}(\mathcal{C}_{i^{'}})- \bar{\mathbf{x}}(\mathcal{C}_i))^{\top} \\
&-A(\mathbf{T})(\bar{\mathbf{x}}(\mathbf{T})-\bar{\mathbf{x}}(\mathcal{C}_i))(\bar{\mathbf{x}}(\mathbf{T})- \bar{\mathbf{x}}(\mathcal{C}_i))^{\top}. \\
\mathbf{U}(\mathcal{C}_{j^{'}}) &= \mathbf{U}(\mathcal{C}_j) + \mathbf{U}(\mathbf{T}) \\
&+A(\mathcal{C}_j)\,\left(\bar{\mathbf{x}}(\mathcal{C}_j)-\bar{\mathbf{x}}(\mathcal{C}_{j^{'}})\right)\,\left(\bar{\mathbf{x}}(\mathcal{C}_j)- \bar{\mathbf{x}}(\mathcal{C}_{j^{'}})\right)^{\top} \\
&+A(\mathbf{T})\,\left(\bar{\mathbf{x}}(\mathbf{T})-\bar{\mathbf{x}}(\mathcal{C}_{j^{'}})\right)\,\left(\bar{\mathbf{x}}(\mathbf{T})- \bar{\mathbf{x}}(\mathcal{C}_{j^{'}})\right)^{\top}.
\end{split}
\end{equation*}

\bibliographystyle{IEEEtran}
\bibliography{ref}
\vspace{5mm}
\begin{IEEEbiography}
	[{\includegraphics[width=1in,height=1.25in,clip,keepaspectratio]{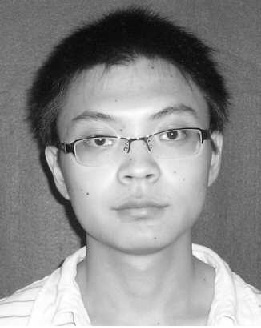}}]{Yiqi Cai}
	is currently a Ph.D candidate in the Department of Computer Science at The University of Texas at Dallas, USA. Before that, he received the bachelor{\rq}s degree in electronic information engineering from The University of Science and Technology of China in 2011, the master's degree in computer science from The University of Texas at Dallas in 2016. His research interests include image segmentation, image registration, 3D reconstruction, GPU algorithms and other related topics.
\end{IEEEbiography}
\begin{IEEEbiography}
	[{\includegraphics[width=1in,height=1.25in,clip,keepaspectratio]{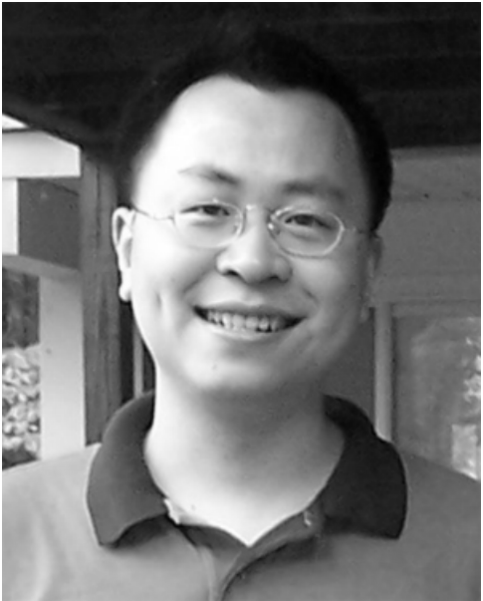}}]{Xiaohu Guo}
	received the PhD degree in computer science from The State University of New York at Stony Brook in 2006. He is an associate professor of computer science at The University of Texas at Dallas. His research interests include computer graphics, animation and visualization, with an emphasis on geometric, and physics based modeling. His current researches at UT Dallas include: spectral geometric analysis, deformable models, centroidal Voronoi tessellation, GPU algorithms, 3D and 4D medical image analysis, etc. He received the prestigious National Science Foundation CAREER Award in 2012.
\end{IEEEbiography}
\begin{IEEEbiography}
	[{\includegraphics[width=1in,height=1.25in,clip,keepaspectratio]{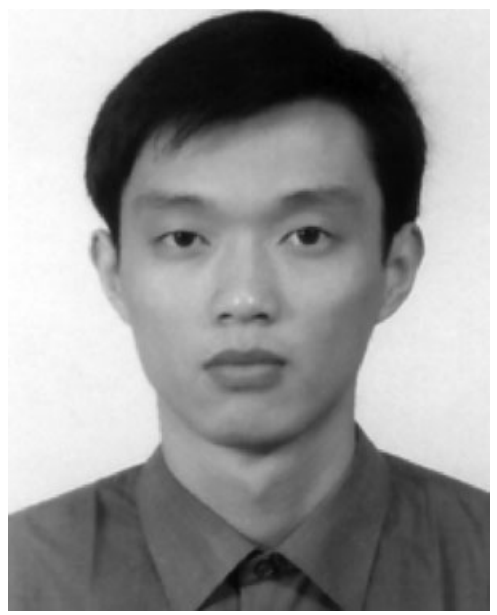}}]{Yang Liu}
	received the bachelor{\rq}s and master{\rq}s degrees in computational mathematics from The University of Science and Technology of China, in
	2000 and 2003, respectively, and the PhD degree in computer science from The University of Hong Kong in 2008. After completing his PhD degree, he worked in Alice group at INRIA/LORIA as a post-doctoral
	researcher starting in 2008. He joined Microsoft Research Asia in 2010. His research interests include geometric modeling, computer-aided geometric design, and architectural geometry.
\end{IEEEbiography}
\begin{IEEEbiography}
	[{\includegraphics[width=1in,height=1.25in,clip,keepaspectratio]{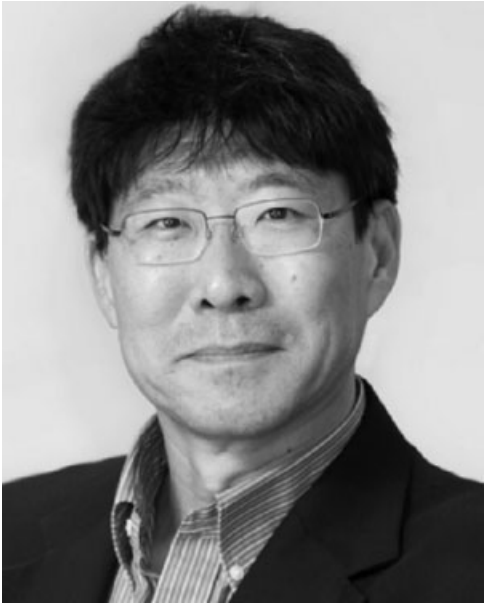}}]{Wenping Wang}
	is a professor of computer science at The University of Hong Kong. His research covers computer graphics, visualization, and geometric computing. He is a journal associate editor of Computer Aided Geometric Design (CAGD), Computers and Graphics (CAG), IEEE Transactions on Visualization and Computer Graphics (TVCG, 2008-2012), Computer Graphics Forum (CGF), and IEEE Computer Graphics and Applications (CG\&A). He has been the program chair of several international conferences, including ACM Symposium on Physical and Solid Modeling (SPM 2006), International Conference on Shape Modeling (SMI 2009), Pacific Graphics 2012, SIAM Conference on Geometric and Physical Modeling 2013 (GD/SPM\rq13), and SIGGRAPH Asia 2013.
\end{IEEEbiography}
\begin{IEEEbiography}
	[{\includegraphics[width=1in,height=1.25in,clip,keepaspectratio]{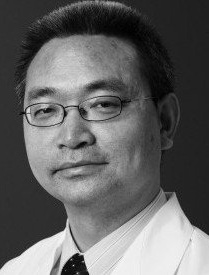}}]{Weihua Mao}
	received his Ph.D. degree in Physics from Peking University in 1999.  He is now a senior staff medical physicist at Henry Ford Hospital, Detroit, MI. His research interests include image-guided radiation therapy, tumor motion management, deformable image registration, and adaptive radiation therapy.
\end{IEEEbiography}
\begin{IEEEbiography}
	[{\includegraphics[width=1in,height=1.25in,clip,keepaspectratio]{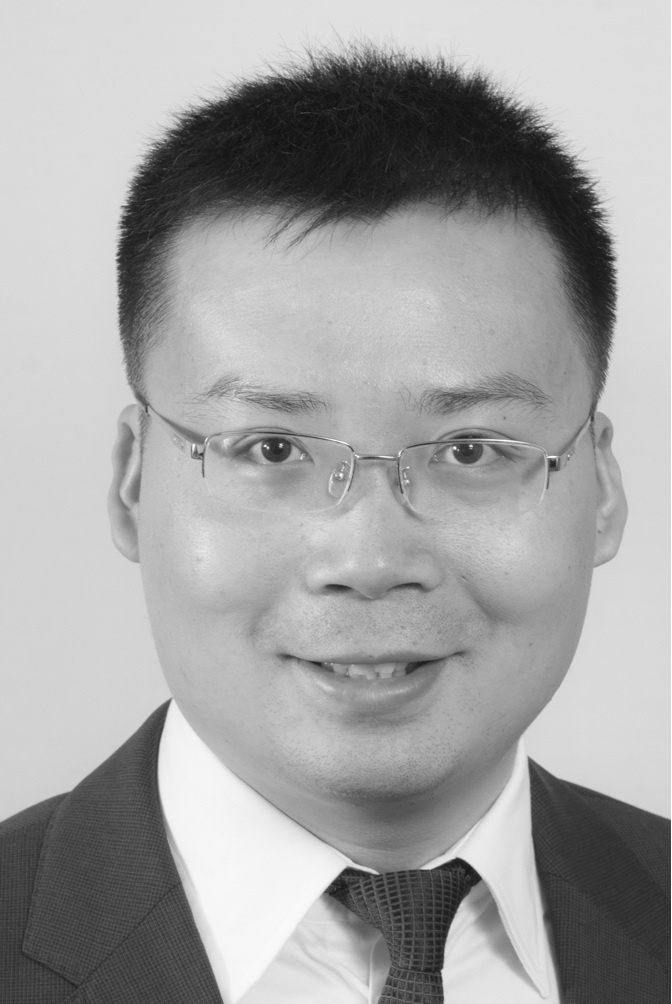}}]{Zichun Zhong}
	received the PhD degree in computer science from the Department of Computer Science at The University of Texas	at Dallas, USA. He is an assistant professor of computer science at Wayne State University. His research interests include computer graphics, geometric modeling (specifically surface and volume mesh generations), medical imaging processing, deformable image registration, image reconstruction, visualization, and GPU algorithms.
\end{IEEEbiography}

\end{document}